\pdfoutput=1
\documentclass{JHEP3}
\usepackage{graphicx,amsmath,amssymb}
\newcommand{\mpl}{M_{Pl}^2}
\newcommand{\rt}{R_{T}}
\newcommand{\rf}{R_F}
\newcommand{\be}{\begin{equation}}
\newcommand{\ee}{\end{equation}}

\newcommand{\vp}{\varphi}

\newcommand{\td}{\text{d}}

\title{Tunneling with negative tension}
\author{Kate Marvel$^1$ and Daniel Wesley$^2$ \\
Department of Applied Mathematics and Theoretical Physics, Cambridge University,  \\
Wilberforce Road, Cambridge CB3 0WA, United Kingdom. \\
$^1$ \email{\tt K.Marvel@damtp.cam.ac.uk}, $^2$ \email{\tt D.H.Wesley@damtp.cam.ac.uk}  }
\date{\today}

\maketitle
\abstract{
We describe a new family of thin-wall instantons, with negative tension bubble walls, that mediate tunneling between Minkowski or de Sitter vacua.  
Some of these instantons can have exponentially enhanced tunneling rates, and would dominate the Euclidean path integral, representing a severe non-perturbative instability in any theory which supports them.
We give two constructions of these instantons in theories which are perturbatively stable, but violate the null energy condition.
One construction uses a scalar field theory with higher-derivative kinetic term, and is similar to the Coleman-de Luccia positive tension instanton.
The other construction employs a negative-tension boundary arising from $Z_2$ orbifolding: it resembles a ``bubble of nothing" which nucleates and grows, consuming the surrounding de Sitter or Minkowski spacetime.
We explain how the spectral flow for fluctutations about the thin-wall tunneling solutions automatically protects causality, for both positive and negative tension instantons.
We comment briefly on the relation of these solutions to a Kalzua-Klein ``bubble of nothing," cosmological models relying on ghost condensates, and string theory orientifolds.
}

\preprint{DAMTP-2008-71}

\begin{document}

\section{Introduction}

In a classic paper \cite{Coleman:1980aw}, Coleman and de Luccia (CdL) introduced a family of instantons in scalar field theory coupled to Einstein gravity. These CdL instantons mediate the decay of metastable de Sitter vacua via bubble nucleation,  and have found numerous applications, especially in cosmology \cite{Brown:1987dd}-\cite{Watson:2006px}.  In this work, we describe a new family of thin-wall negative-tension instantons that are similar to CdL instantons, but cannot be realized in a canonical scalar field theory.  We give two constructions of these new instantons: one using higher-derivative scalar field actions, the other using negative-tension boundaries similar to those which appear in some higher-dimensional constructions.  The first realization is a negative-tension analogue of the classic thin-wall CdL instanton, while the other appears as a hole in spacetime or a ``bubble of nothing" which expands and consumes an entire horizon volume. Some of the new instantons  have an exponentially \emph{enhanced} tunneling rates, and so would dominate the Euclidean gravitational path integral.  We show how the spectrum of fluctuations around the tunneling solutions automatically protects causality, quenching the tunneling rate for acausal solutions.  This causality protection works for both positive and negative tension instantons, and sheds light on the delicate issue of negative modes in CdL tunneling 
\cite{Coleman:1987rm}-\cite{ColemanSteinhardt}.  The sub-family of new instantons with exponentially enhanced tunneling rates represent a catastrophic nonperturbative quantum instability in the theories that support them, theories which can otherwise be perturbatively stable.  These new tunneling solutions therefore provide an important consistency constraint on a variety of models which employ negative-tension boundaries or higher-derivative scalar field actions. 

 In the thin-wall limit, the Euclidean CdL instanton is two $S^4$ of different radii (representing the true and false de Sitter vacua) joined along an $S^3$ (representing the bubble wall).  The Israel matching conditions determine the tension of the bubble wall in terms of the two de Sitter cosmological constants and the bubble radius.  Depending on the specific configuration chosen, the required tension can be either positive or negative.  Classic CdL instantons have only positive-tension bubble walls.  Normally the negative-tension solutions are discarded as unphysical, because a canonical scalar field with a potential can be used to construct walls of positive tension only.  However, as we show in this paper, if one allows for the possibility of negative-tension bubble walls, then it is possible to construct a new family of tunneling instantons.

There are several good reasons to be suspicious of these new CdL-type solutions with negative-tension walls, but we argue that they make for perfectly sensible instantons.  There are two essential features that must be checked before an instanton can be classed as a tunneling solution. The first requirement is the existence of a finite-action solution to the Euclidean equations of motion.  This requirement is easily satisfied by joining two $S^4$ along an appropriate $S^3$, in the same way as for the traditional thin-wall CdL instantons.  The other requirement is that the action for fluctuations about this Euclidean solution be a quadratic form with precisely one negative eigenvalue.\footnote{An odd number of negative eigenvalues may also be permitted; see eg \cite{Gratton:2000fj},\cite{Hackworth:2004xb}.}  The corresponding eigenfunction or ``negative mode" ensures that the ground state energy acquires an imaginary part, which is required if the instanton is to describe the decay of a false vacuum.  Without a negative mode, the instanton merely corrects the ground-state energy of the system.  We show that some of the negative-tension instantons have the required negative mode.  Therefore our negative-tension instantons possess the two features needed to mediate vacuum decay, and can be regarded as valid tunneling solutions.

Having established that the negative-tension solutions are valid instantons, we provide two concrete realizations of them.  The negative-tension instantons cannot be constructed using a canonical scalar field with potential, as the traditional CdL instantons are, since a canonical scalar field can only give rise to a positive tension bubble wall.  As we show below, it is possible to construct these instantons if the scalar field theory has a higher-derivative kinetic term.   As with all higher-derivative theories, it is important to check that there are no ghostlike (negative norm) fluctuations about the background solution.  Using a simple higher-derivative scalar Lagrangian, we construct an explicit ghost-free, negative-tension bubble wall solution that asymptotes to Minkowski space.  This gives an example of a higher-derivative theory that is perturbatively stable and ghost-free, but which possesses a nonperturbative instability to the formation of negative-tension CdL bubbles.

Our second construction uses ingredients similar to those which appear in some higher-dimensional constructions.  Specifically, we consider theories in which stress-energy can be localized on boundaries of spacetime.  One example is the orientifold in string theory, a combined spacetime and worldsheet discrete transformation which creates a boundary of spacetime that typically carries negative  tension \cite{Polchinski}.  In supergravity constructions where the compactification space has boundaries, global constraints often require one (or more) of the boundaries to carry negative tension (eg, \cite{Bergshoeff:2000zn,Lukas:1998yy}).  These negative-tension boundaries are also useful in warped compactifications, such as the $S^1/Z_2$ Randall-Sundrum construction \cite{Randall:1999ee,Randall:1999vf} and flux compactifications \cite{Verlinde:1999fy}-\cite{Burgess:2006mn}.  In all these cases, the boundary is realized as the fixed point locus of a discrete spacetime transformation\footnote{In string models this can be combined with a worldsheet parity transformation (orientifolding).} (orbifolding).  This orbifolding is essential for the perturbative stability of the boundaries, for it truncates boundary fluctuation modes which would otherwise grow catastrophically.
For our purposes, we simply assume that the action for the theory allows for boundaries of spacetime which carry positive or negative tension, and that these boundaries are realized as $Z_2$ orbifoldings.

With these assumptions, we show there exists an instanton describing the appearance of a ``hole" in de Sitter or Minkowski space.  The boundary of the hole carries negative tension, and the spacetime solution is obtained by a simple $Z_2$ orbifolding of a suitable negative-tension thin-wall instanton solution.  Thus, locally, the instanton is constructed in the same way as negative-tension boundaries arising in the scenarios previously mentioned.  Inside the hole there is nothing -- no spacetime and no matter fields.  The hole grows at a speed approaching that of light, and eventually consumes an entire horizon volume, or collides with another hole.  While these solutions may appear bizarre, they are not without precedent. Examples include the ``bubble of nothing" described by Witten in the context of five-dimensional Kaluza-Klein theory \cite{Witten:1981gj} and later examples in string \cite{Aharony:2002cx} and M theory \cite{Fabinger:2000jd,Horava:2007hg,Horava:2007yh}.  There are no positive-tension versions of the $Z_2$ instantons we describe, so they are a unique feature of theories which allow boundaries with negative tension.

Normally we would be safe in ignoring negative tension solutions because they seem to require exotic matter actions with higher derivatives, or strange objects with negative tension.  However,  many current ideas in fundamental physics and cosmology are driving us to include precisely these sorts of features.  One commonality shared by any theory which supports negative-tension instantons is violation of the null energy condition (NEC).  The NEC requires that, for any null vector $n^\mu$, the stress energy tensor $T_{\mu\nu}$ satisfies \cite{HawkingEllis}
\be
T_{\mu\nu} n^\mu n^\nu \ge 0
\ee
It is the weakest of the classic energy conditions.  Violation of the NEC is often associated with various pathologies \cite{Tipler:1976bi}-\cite{Buniy:2006xf}, solutions of the Einstein equations with strange properties \cite{Morris:1988cz}-\cite{Caldwell:2003vq}, and violations of gravitational thermodynamics \cite{Rubakov:2004eb}-\cite{Eling:2007qd}.  It seems likely that the instabilities mediated by our new instantons are related to their violation of this condition. Nonetheless NEC violation is incorporated in a variety of models currently under discussion.  Inflationary and de Sitter string constructions often include NEC violating sources such as orientifolds \cite{Verlinde:1999fy}-\cite{Burgess:2006mn}. Some alternatives to inflation with a big crunch/big bang transition, including new ekpyrosis \cite{Buchbinder:2007ad}, the ``breathing Universe" \cite{Guendelman:2008ys} and other models \cite{Creminelli:2006xe}-\cite{Cornalba:2002nv}, also rely on NEC violation.  Negative tension boundaries may also arise in scalar field theories with nonminimal coupling \cite{Lee:2006vka,Lee:2007dh}.  A broad variety of models with extra dimensions must violate the NEC in order to produce cosmic acceleration,as required for inflation or the current epoch of dark-energy dominated expansion \cite{Wesley:2008fg,Wesley:2008de,SteinhardtWesley}.  herefore it is very important to know if and how the NEC can be consistently violated.  Our two constructions of negative-tension CdL instantons employ two means of NEC violation (ghost condensates and negative-tension boundaries) which are perturbatively stable, but which can have a catastrophic nonperturbative instability in the form of negative-tension bubble nucleation.  This is consistent with the intuition that NEC violation usually leads to trouble.  The arguments we present here certainly do not indicate that all theories with ghost condensates or negative tension boundaries have problems.  But, given how essential such NEC-violating elements seem to be in many important constructions, the instabilities mediated by these new instanton solutions do suggest that a deeper understanding of these NEC-violating elements would be useful.  Ensuring that the solutions we describe here are absent provides a new consistency check on models that 
allow NEC violation.

There is an important point regarding causality which must be established for the new negative-tension instantons.  This issue, related to the possibility of superhorizon bubble nucleation, has remained a puzzling feature even for thin-wall CdL instantons.  The nucleation (``bounce") radius $r_{\rm nuc}$ of the CdL bubble depends on the cosmological constants in the true and false vacuum, as well as the potential for the scalar field, but there are ranges of these parameters for which the bubble nuclation radius is larger than the false vacuum de Sitter horizon $R_{\rm dS}$.  Taken literally, the nucleation of such a super-horizon bubble would be acausal.  Some authors  have circumvented this issue by confining their analysis to the causal diamond of de Sitter space \cite{Brown:2007sd}, while others have recognized the existence of superhorizon solutions and discarded them as unphysical \cite{ColemanSteinhardt}, \cite{Maloney:2002rr}- \cite{Charmousis:2008ce}.  This issue is easy to ignore for positive tension bubbles: the tunneling rate $\Gamma$ is given by
\be\label{e:CdLrate}
\Gamma = A e^{-B}
\ee
and a naive estimate of $B$  indicates that the rate for nucleating positive-tension bubbles at nearly the horizon radius is very small.  For negative tension bubbles, an estimate for $B$ suggests that the rate for horizon-size bubble nucleation can be quite large, forcing us to confront the causality issue.

We show that thin-wall instantons automatically protect causality,  by studying the action for small fluctuations around the thin-wall instanton solution.  These fluctuations enter into the functional determinant prefactor $A$ in the WKB tunneling rate expression (\ref{e:CdLrate}). There is a fluctuation mode whose action switches sign as the nucleation radius crosses the horizon radius.  We argue that this changes the character of the Euclidean solution from a tunneling solution (for $r_{\rm nuc} < R_{\rm dS}$) to a ground-state energy correction (for $r_{\rm nuc}> R_{\rm dS}$).  The super-horizon bubble then represents a virtual process, and no more violates causality than does a virtual particle propagating outside the light cone in quantum field theory.  This switch in behavior is driven by an ``enhanced symmetry point" (ESP) when the nucleation radius equals the horizon radius, at which the Euclidean action for a particular fluctuation mode vanishes.  The ESP implies that the prefactor $A$ of the semiclassical tunneling rate formula blows up as $r_{\rm nuc} \to R_{\rm dS}$, indicating that the semiclassical approximation for the tunneling rate should not be trusted for nearly horizon-size bubbles.  Away from this divergence, where the semiclassical calculation can be trusted, the absence of a negative mode for superhorizon bubble nucleation automatically protects causality.

This paper is organized as follows. In Section \ref{s:review}, we review the basics of thin-wall CdL tunneling and describe the new thin-wall negative-tension solutions in parallel.  We establish the existence of solutions to the Euclidean equations of motion with both positive and negative tension, and describe which configurations possess the requisite negative mode.  In Section \ref{s:ghost} we construct a negative-tension instanton solution using a specific higher-derivative scalar field Lagrangian.  In Section \ref{s:orientifold} we describe a realization of the negative-tension solutions in terms of negative-tension $Z_2$ orbifold boundaries.  We describe the resulting ``bubble of nothing," and discuss its relationship to other similar solutions.  Section \ref{s:interpretation} is devoted to the problem of super-horizon bubbles.  We give our interpretation of the quenching of the tunneling rate when the bubble nucleation radius exceeds the de Sitter horizon.  Our conclusions and summary are given in Section \ref{s:conclusions}.

Throughout this work we use units in which $8\pi G = 1/M_{pl}^2$, and we define the cosmological constant $\Lambda$ in terms of the vacuum energy density $\rho_{\rm vac}$ by $\Lambda = \rho_{\rm vac} / M_{pl}^2$, so $\Lambda$ has units of $M_{pl}^2$.

\section{Generalizations of thin-wall instantons}\label{s:review}

In this section we establish the existence of thin-wall tunneling solutions with negative-tension bubble walls.  We combine this with a review of the basics of thin-wall CdL tunneling.  In Section \ref{ss:FourBubbles} we present the four classes of thin-wall bubble solutions, and show that two of them correspond to negative-tension bubbles, and compute the tunneling rates for these new negative-tension solutions.  We discuss the negative mode condition necessary for a Euclidean solution to describe tunneling in Section \ref{ss:NegModes}.  

\begin{figure}
\begin{center}
\includegraphics[scale=.6]{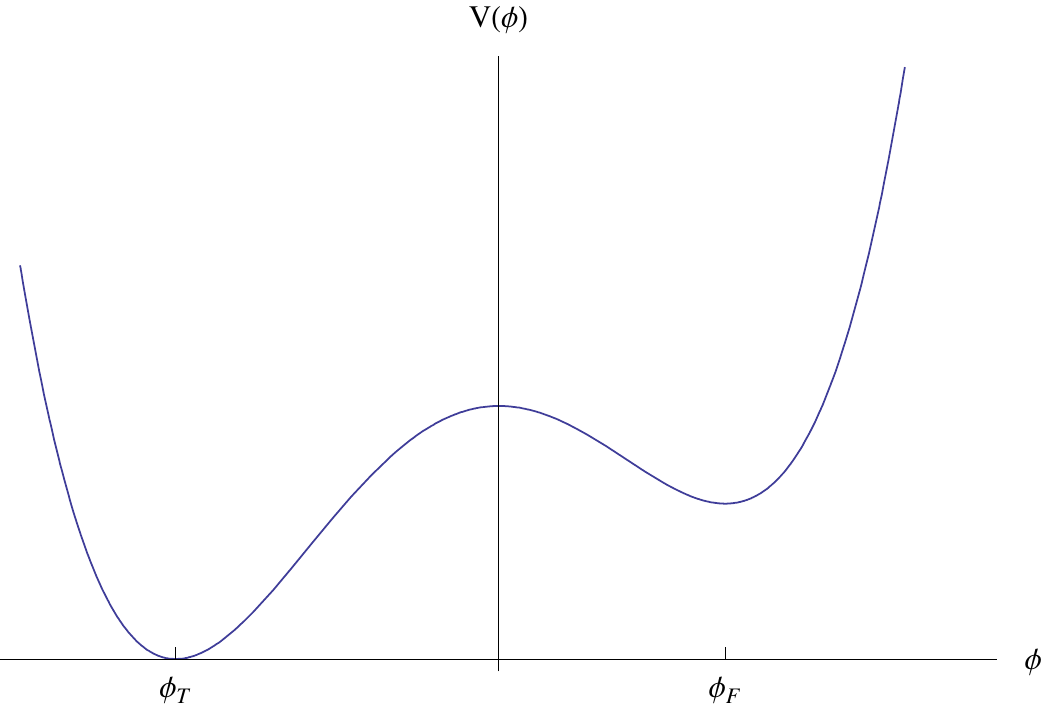}
\caption{Coleman-DeLuccia potential.  The difference in potential energy between the minima is $\epsilon$.  The bubble wall tension is related to the potential barrier between the two minima.}
\label{CDpotential}
\end{center}
\end{figure}

The original work on bubble nucleation studied tunneling between the two minima in a scalar field theory coupled to Einstein gravity \cite{Coleman:1977py,Callan:1977pt,Coleman:1980aw}.  It was found that the problem of gravitational tunneling simplified in the thin-wall approximation, in which the vacuum separation $\epsilon=\Lambda_F-\Lambda_T$  is negligible compared to the height of the potential barrier between the minima (See Figure \ref{CDpotential}).  Tunneling is described by the appearance of a ``bubble" of true vacuum within the surrounding false vacuum space.  The size of the bubble at nucleation and its subsequent trajectory are determined by the parameters $\epsilon$ and the effective bubble wall tension $T$.  The latter can be computed from the potential for the scalar field.  

In this paper we work entirely with thin-wall bubbles.  This approximation is valid for a wide range of models which rely on a potential, and for all models in which vacuum decay is mediated by branes.   We consider the decay of de Sitter space with positive cosmological constant $\Lambda_F$ to a space with positive cosmological constant $\Lambda_T$, where $\Lambda_T \le \Lambda_F$.  The boundary between the two spaces is assumed to be infinitely thin and described purely by its tension $T$.  We are not necessarily assuming that there is an underlying scalar field theory, and indeed some of the bubbles we introduce cannot be described by a canonical scalar field.  There are three main benefits to our approach.   The first is that it admits a simple parameterization of the instanton solution: the tunneling problem is entirely specified by the wall tension and Euclidean de Sitter radii $R_{(F,T)}=\left(3/\Lambda_{(F,T)}\right)^{1/2}.$  Second, large tensions are difficult to incorporate in the traditional CdL picture, because the thin-wall approximation may not be valid for the potentials that produce large-tension bubbles.  Third,  our method allows us to treat $T$ as a free parameter, which simplifies the generalization to negative $T$, which cannot be realized in the traditional CdL family of solutions.  

\subsection{Four bubbles}\label{ss:FourBubbles}

We begin by analytically continuing to Euclidean time, in which the true and false vacua both have the geometry of four-spheres. In this picture, by ``gluing"  sections of these manifolds along a boundary, we can satisfy the requirement that the solution asymptotically approach the false vacuum in the past and the true vacuum in the future.

\begin{figure}
\begin{center}
\includegraphics[scale=.6]{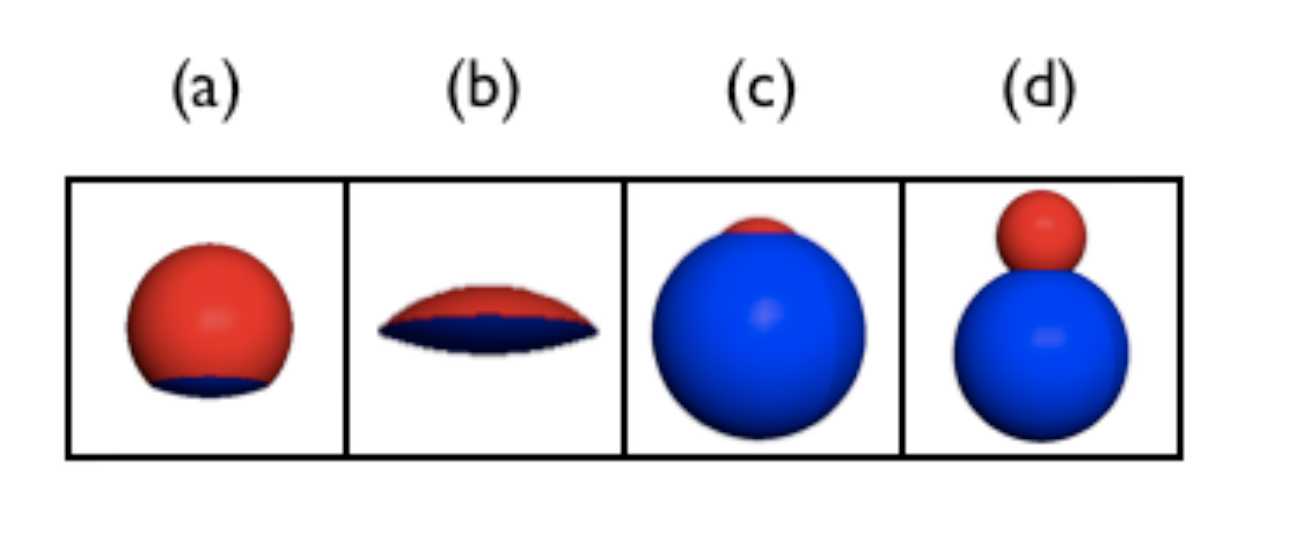}
\caption{Four types of instanton.  Red represents false vacuum, blue true.  At the ``north pole" (top of each diagram) $\sigma=0$, and at the ``south pole" (bottom of each diagram) $\sigma = \pi R_F$.  The coordinate $\sigma$ is discontinuous across the join. The two configurations at the top correspond to the choice $\sigma_{T}/\rt > \pi /2$.  The bottom configurations correspond to  $\sigma_{T}/\rt < \pi /2$.}  \label{fourclasses}
\end{center}
\end{figure}

The instanton metric $\tilde{g_{\mu \nu}}$  is given by \cite{Coleman:1980aw}
\be
ds^2 =\tilde{g_{\mu \nu}}\td x^{\mu} \td x^{\nu}= \td \sigma^2 +b^2(\sigma) \td \Omega_{(3)}^2\label{EucMetric}
\ee
with $\td \Omega_{(3)}^2$ the metric on the unit three-sphere $S^3$.  Radial distances are measured by $\sigma$, which is defined so that $\sigma=0$ at the ``north pole" (false vacuum) and $\sigma=\pi R_T$ at the ``south pole" (true vacuum).  In the Euclidean picture, the bubble wall is the boundary manifold separating false vacuum from true.  In the thin-wall approximation, it is a round $S^3$ and its radius $b(\sigma)$ is given by
\be
b(\sigma) = \left\{ \begin{array}{l}
\rf \sin(\sigma/\rf) \; \; \mathrm{if}\; \; 0 \leq \sigma \leq \sigma_{wall} \\
\rt \sin(\sigma/\rt)\; \; \mathrm{if}\; \; \sigma_T\leq \sigma \leq \pi \rt
\end{array}\right. \label{metric}
\ee
where $\sigma_{wall}$ and $\sigma_T$ are the coordinates of the boundary surface, or "join", in the false and true vacuum coordinate patches, respectively.  The coordinate $\sigma$ is discontinuous across the join, but this is a coordinate discontinuity reflecting the two coordinate patches we have used to cover the Euclidean spacetime.  The wall coordinates in either patch are constrained by demanding the metric be continuous across the join, which requires
\be\label{e:AngleContinuous}
\rt \sin(\sigma_{T}/\rt)=\rf \sin(\sigma_{wall}/\rf).
\ee
The equation (\ref{e:AngleContinuous}) has four solutions for $(\sigma_T,\sigma_{wall})$ since it is invariant under the transformations 
\be\label{e:SigmaTX}
\sigma_T \to \pi R_T - \sigma_T \qquad \sigma_F \to \pi R_F - \sigma_F
\ee
where either transformation acts independently.  This means that there are four ways to ``glue" together the true and false vacua across the join,\footnote{These configurations complement other enumerations, such as the classification of \cite{Blau:1986cw,Aguirre:2005xs,Aguirre:2005nt}, in which the tension is restricted to be positive, but the metric is generalized to the Schwarzschild-de Sitter form.},  as illustrated in Figure \ref{fourclasses}.  The join coordinates determine the Euclidean ``join radius" $\rho$ by
\be\label{e:RhoFunSigma}
\rho = b(\sigma_{wall}) = \rf  \sin(\sigma_{wall}/\rf)= \rt \sin(\sigma_{T}/\rt)
\ee
which is also invariant under the transformations (\ref{e:SigmaTX}).
The join radius is not the physical radius of the bubble wall as measured in either the true or false Euclidean vacuum spacetime regions, but the radius of the $S^3$ in the embedding Euclidean space.

The join angle $\sigma_{wall}$ is fixed by minimizing the Euclidean action for the instanton described by the metric $\tilde{g}$:
\be\label{einsteinhilbert}
S_E=-\frac{\mpl}{2} \int \left(\tilde{R} -2\tilde{\Lambda}\right) \, \sqrt{\tilde{g}}\, \td^4 x. + T \int \sqrt{h} \, \td^3 y - \int K \sqrt{h} \, \td^3 y
\ee    
Here $y$ are the coordinates, $h$ the induced metric, $K$ the extrinsic curvature on the wall, and $\tilde{R}$ the Ricci scalar for the instanton manifold, with 
\be
\tilde{\Lambda} = \left\{ \begin{array}{l}
3 \rf^{-2} \; \; \mathrm{in}\;  \mathrm{the} \; \mathrm{false}\;  \mathrm{vacuum} \\
3 \rt ^{-2}\; \; \mathrm{in}\;  \mathrm{the} \; \mathrm{true}\;  \mathrm{vacuum} 
\end{array}\right. \label{metric}
\ee
For the $O(4)$-symmetric metric (\ref{EucMetric}), the action can be split into a volume contribution from each vacuum and a surface energy contribution from the boundary wall.  After  integration by parts we have
\begin{eqnarray}
S_E=T\int \delta(\sigma-\sigma_{wall})\td\sigma \; \td\Omega_{(3)}^2-\frac{\mpl}{2} \int_{false} 6(b+b b'^2 - b^3 \rf^{-2}) \td\sigma \; \td\Omega_{(3)}^2 \nonumber \\ -\frac{\mpl}{2}
 \int_{true}  6(b+b b'^2 - b^3 \rt^{-2}) \td\sigma \; \td\Omega_{(3)}^2. \label{prettyaction}
\end{eqnarray}
The action for fixed $(\rt, \rf, T)$ depends only on the join radius and can be written as 
\be
\frac{S_E}{2 \pi^2} = T \rho^3 - 2\mpl \rf^2\left\{ \mathcal{F} \left(1-\frac{\rho^2}{\rf^2}\right)^{3/2} +1\right\}- 2\mpl \rt^2\left\{ \mathcal{T} \left(1-\frac{\rho^2}{\rt^2}\right)^{3/2}+1\right\}.
\label{prettyac}
\ee
The variables $\mathcal{T}$ and $\mathcal{F}$ take the values $\pm 1$ and measure the gravitational volume contributions of the true and false vacuum manifolds respectively.   Each instanton configuration in Figure \ref{fourclasses} can then be labeled by the pair $(\mathcal{F},\mathcal{T})$, as shown in Table \ref{sumtable}.
For any choice of $(\mathcal{F},\mathcal{T})$, the action is stationary at the nucleation radius\footnote{It is also stationary at $\rho=0$, corresponding to the trivial configuration of false vacuum with no bubble.}
\be
\rho=\frac{4 \mpl \rf^2 \rt^2 |T| }{\sqrt{16 M_{Pl}^8(\rf^2-\rt^2)^2 + 8 M_{Pl}^4 \rt^2 \rf^2 (\rf^2 + \rt^2)T^2 + \rf^4 \rt^4 T^4}} .
 \label{e:NucRadius}
\ee
This single join radius corresponds to four different possible instanton configurations, labeled by the four choices for $(\mathcal{F},\mathcal{T})$.

\begin{table}
\begin{center}
\begin{tabular}{|c| c|  c | c | c | c | }
\hline
Type & $\mathcal{F}$ & $\mathcal{T}$ &  bubble tension & bubble size & negative mode? \\ \hline
\hline
(a) & $+1$& $-1$& small, positive & sub-horizon & yes \\ \hline
(b) & $-1$&$-1$& large, positive & super-horizon & no \\ \hline
(c) &$-1$& $+1$ & small, negative & super-horizon & no \\ \hline
(d) &$+1$ & $+1$ & large, negative & sub-horizon & yes \\ \hline
\end{tabular}
\caption{A summary of the properties of the four instanton configurations.}\label{sumtable}
\end{center}
\end{table}

The expression (\ref{e:NucRadius}) gives a nonzero nucleation radius for both traditional positive-tension, as well as negative-tension, bubble walls.  To ensure that the solution for negative tension is not a spurious root, we check that the Israel matching conditions are satisfied. This determines the wall tension in terms of $(R_F,R_T)$ and the discrete parameters $(\mathcal{F},\mathcal{T})$.
In the thin-wall approximation, the boundary is a three-sphere whose metric $h_{ab}$ scales with the join radius $\rho$:
$$h_{ab} \td x^a \td x^b = \rho^2  \td \Omega_{(3)}^2 .$$
Its stress-energy can be written in the form 
\be
T_{ab} = -T h_{ab},
\ee
where $T$ is the tension on the boundary.  With this choice of coordinates, the vector $\xi^{\mu}$ normal to the boundary wall points in the $\hat{\sigma}$ direction, and the extrinsic curvature induced on the boundary from either side is therefore
\be
K_{ab}= \nabla_a \xi_b = -\Gamma^{\sigma}_{ab}=\frac{b'}{b} h_{ab}. \label{exdef}
\ee
The jump in extrinsic curvature across the boundary is related to its three-dimensional surface energy  by the Israel matching condition
\be
K_{ab}^F-K_{ab}^{T} = -\frac{(T_{ab}-\frac{1}{2} h_{ab} T)}{\mpl}
\ee
which reduces to the simple constraint
\be\label{e:TensionRFRT}
-\frac{T}{2\mpl} = 
\mathcal{F}\sqrt{\frac{1}{\rho^2}-\frac{1}{\rf^2}}+\mathcal{T}\sqrt{\frac{1}{\rho^2}-\frac{1}{\rt^2}}
\ee
Since $R_F \le R_T$ the absolute value of the first term on the left hand side is never larger than the second.  Thus the sign of $\mathcal{T}$ completely determines the sign of the tension $T$. When $\mathcal{T} = -1$ the tension is positive, and when $\mathcal{T} = +1$ the tension is negative.  We conclude that the configurations (a) and (b) describe tunneling mediated by positive-tension bubble walls, directly analogous to the CdL instanton.  Types (c) and (d) correspond to tunneling with negative-tension bubbles.

\begin{figure}
\begin{center}
\includegraphics[scale=.6]{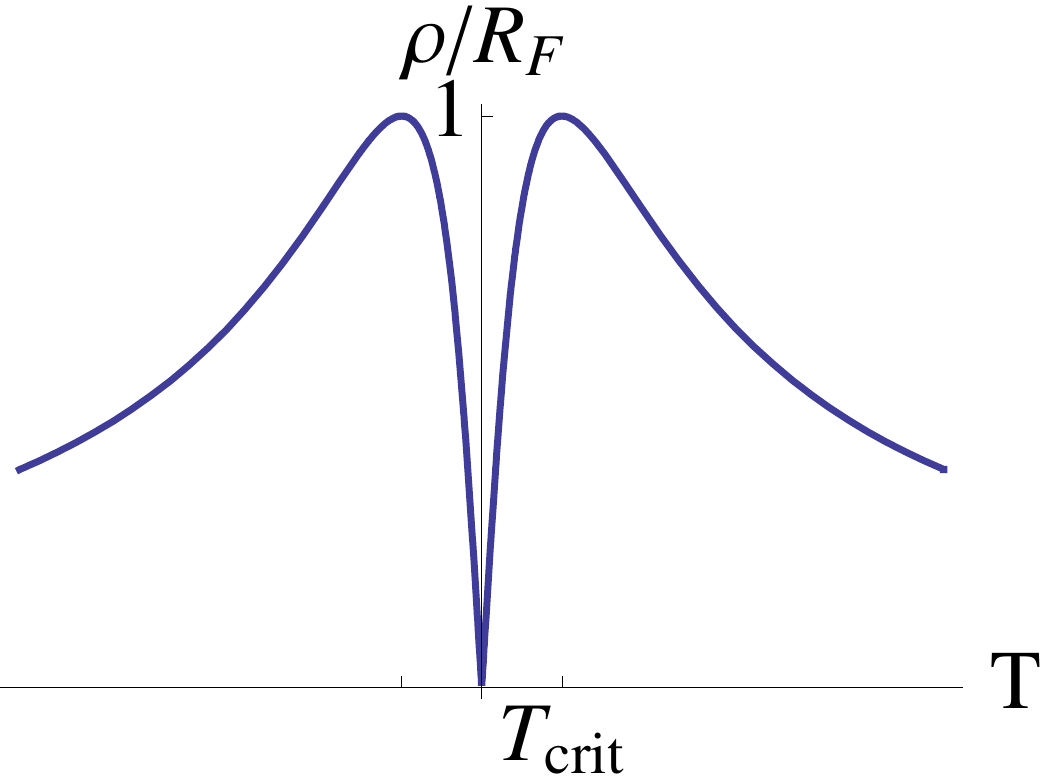}
\caption{Bubble nucleation radius as a function of wall tension (fixed cosmological constants).  When $T=\pm T_{crit}$ then the nucleation radius $\rho = R_F$, the false vacuum horizon.  For positive tensions, $T > T_{crit}$ corresponds to super-horizon bubble nucleation and $T < T_{crit}$ corresponds to sub-horizon bubble nucleation.  For negative tensions, the situation is reversed: $|T| > |T_{crit}|$ corresponds to sub-horizon bubble nucleation while $|T| < |T_{crit}|$ corresponds to super-horizon bubble nucleation. }
\label{rhoT}
\end{center}
\end{figure}

The relation (\ref{e:RhoFunSigma}) gives two proper bubble radii $\sigma_{wall}$ for each choice of $\rho$.  One of these radii is larger than the false vacuum horizon ($\sigma_{wall} > R_F$) and the other is smaller ($\sigma_{wall} < R_F$).  Each solution for fixed $\rho$ corresponds to a different choice of $\mathcal{F}$ and hence to a different tension by (\ref{e:TensionRFRT}) as illustrated in Figure \ref{rhoT}.  Instantons of Type (a), with $\mathcal{F} = -1$ have small, positive tensions.  Since $\mathcal{F} = -1$ there is a small amount of true vacuum patched onto a larger portion of false.  Therefore these instantons nucleate at sub-horizon size ($\sigma_{wall} < R_F$).  For increasing positive tension, the join radius approaches the equator of the false vacuum sphere, reaching it at the critical tension
\be\label{e:Tcrit}
T_{crit}= 2 \mpl \sqrt{\frac{1}{\rf^2}-\frac{1}{\rt^2}},
\ee
as illustrated in Figure \ref{rhoT}.    If the tension is positive and its magnitude exceeds this critical value, the instanton is of Type (b) and the volume of the false vacuum semisphere decreases as the tension increases in magnitude.   Type (b) instantons nucleate at super-horizon sizes ($\sigma_{wall} > R_F$).

 The negative tension case is similar, but with some signs reversed.  The $\mathcal{F}=-1$ negative-tension instantons still correspond to super-horizon bubbles, but because $\mathcal{T}$ has switched sign, these now correspond to small-magnitude negative tensions.  Therefore, for small negative tensions the instanton is of Type (c).  For negative tensions of increasing magnitude, the instanton solution contains a smaller and smaller portion of false vacuum.  When the tension reaches the negative of the critical value (\ref{e:Tcrit}) the join radius once again coincides with the horizon of the false vacuum.  Large-magnitude negative tensions therefore produce instantons of Type (d) with bubbles that nucleate within the false-vacuum horizon ($\sigma_{wall} < R_F$).  We return to these issues in more detail in Section \ref{s:interpretation} where we study the causal properties of the instanton solutions.

Having calculated an explicit form for the Euclidean action, we can then write the decay rate per unit volume as 
\be\label{e:TunnelExpression}
\Gamma =A\exp(-B)
\ee
where 
\be\label{e:Bprescription}
B=S_E-S_{stay}.
\ee  
The Euclidean action $S_E$ is given by (\ref{prettyaction}) evaluated at the join radius $\rho$. The normalizing factor $S_{stay}$ is the Euclidean action for the initial configuration of pure false-vacuum de Sitter, giving
\be
S_{stay}=\lim_{T \rightarrow 0}S_E=  -8 \pi^2 \mpl \rf^2.
\ee
For positive-tension tunneling, analogous to the CdL tunneling instanton, the exponential factor $B$ in the tunneling  rate is then given by
\be
\frac{B_+}{48 \pi^2 \mpl}=\frac{3 T^2 \Lambda_{+} + 4 M_{pl}^4\Lambda_{-}^2-\Lambda_-\sqrt{9 T^4 + 16M_{pl}^8\Lambda_{-}^2+24 M_{pl}^4 T^2 \Lambda_{+}}}{(\Lambda_+^2-\Lambda_-^2)\sqrt{9 T^4 + 16 M_{pl}^8 \Lambda_-^2+24 M_{pl}^4 T^2 \Lambda_{+}}}
\ee
where we have set $\Lambda_{+}=\Lambda_F+\Lambda_T$ and $\Lambda_{-}=\Lambda_F-\Lambda_T$.  In the limit of small difference between the cosmological constants, we have $\Lambda_T = \Lambda$ and $\Lambda_F = \Lambda +\epsilon$ in which case $B_+$ becomes
\be\label{e:BplusEpsilon}
B_+ = +\frac{12\pi^2 M_{pl}^2 \sqrt{\epsilon}}{\Lambda^{3/2}} \left(\frac{T}{T_{crit}}\right) + \mathcal{O}(\epsilon)
\ee
When the cosmological constant $\Lambda$ is sub-Planckian, $\Lambda \ll M_{pl}^{-2}$, then $B_+$ is very large and positive, and the tunneling rate is highly suppressed.  The tunneling rate can be enhanced if the cosmological constants are very close $\epsilon / \Lambda \ll 1$, or if if the tension is small $T \ll T_{crit}$.

The story is slightly different for the instanton with negative-tension walls.  The tunneling rate expression is
\be
\frac{B_-}{48 \pi^2 \mpl}=-\frac{3 T^2 \Lambda_{+} +4 M_{pl}^4\Lambda_{-}^2+\Lambda_-\sqrt{9 T^4 + 16M_{pl}^8\Lambda_{-}^2+24 M_{pl}^4 T^2 \Lambda_{+}}}{(\Lambda_+^2-\Lambda_-^2)\sqrt{9 T^4 + 16 M_{pl}^8 \Lambda_-^2+24 M_{pl}^4 T^2 \Lambda_{+}}}
\ee
In the limit of very small separation $\epsilon$ between the cosmological constants, this reduces to
\be\label{e:BminusEpsilon}
B_- = -\frac{12\pi^2 M_{pl}^2 \sqrt{\epsilon}}{\Lambda^{3/2}} \left(\frac{|T|}{T_{crit}}\right) + \mathcal{O}(\epsilon)
\ee
There are several crucial differences between (\ref{e:BplusEpsilon}) and (\ref{e:BminusEpsilon}).  The most important is that the sign of $B_-$ is has switched, so that tunneling is exponentially enhanced, instead of suppressed as in the CdL case (\ref{e:BplusEpsilon}).  A further enhancement arises from the tensions of allowed instantons.  As we describe in more detail in Section \ref{ss:NegModes}, only negative-tension bubbles with $|T| > T_{crit}$ describe valid tunneling instantons.  That means that the $|T|/T_{crit}$ factor in (\ref{e:BminusEpsilon}) serves to further enhance the tunneling rate, since it must be larger than unity.

\subsection{Negative modes} \label{ss:NegModes}

Small fluctuations about the instanton solution determine both the prefactor $A$ in (\ref{e:TunnelExpression}) and whether we can interpret the solution as a tunneling solution \cite{Callan:1977pt,Coleman:1977py,Coleman:1980aw},\cite{Coleman:1987rm}-\cite{ColemanSteinhardt}.  Denoting the various fields that may be present in the action (scalar fields, metric, etc) by $\varphi$, and by $\varphi_0$ the background solution for the instanton, then expanding the Euclidean action about the instanton solution as $\varphi = \varphi_0 + \delta\varphi$ gives a quadratic term of the schematic form\footnote{Lorentz and other indices on $\varphi$ and $\mathcal{D}_E$ have been suppressed for clarity.} 
\be
S_E = \cdots + \int \delta \varphi \, \mathcal{D}_E \, \delta  \varphi \, \sqrt{g} \, \td^4 x + \cdots
\ee
where we have introduced the operator $\mathcal{D}_E$.The prefactor $A$ is related to the functional determinant of $\mathcal{D}_E$ through 
\be
A \propto \left( \text{det}' \,\mathcal{D}_E \right)^{-1/2}
\ee
where $\text{det}'$ indicates that zero eigenvalues of $\mathcal{D}_E$ are omitted.  Of more importance for our present purposes, the eigenvalues of $\mathcal{D}_E$ determine what type of instanton we have -- whether it can represent a tunneling process, or is a virtual process which merely corrects the ground-state energy of the system.  To be a tunneling solution, $\mathcal{D}_E$ must have a negative eigenvalue.  Then the instanton marks a saddle point of the action and there exists at least one direction in which fluctuations may decrease the action.  The instanton is then the leading correction to the imaginary part of the energy and its existence renders the false vacuum state unstable.  To have a true tunneling instanton, we must demonstrate the existence of such a negative mode.

We study the $O(4)$-invariant mode corresponding to fluctuations in the Euclidean bubble wall radius $\rho$.  The full spectrum of eigenvalues for $\mathcal{D}_E$ in CdL tunneling has remained a challenging and controversial problem \cite{Coleman:1987rm}-\cite{ColemanSteinhardt}.  Nonetheless in various examples showing the existence of an $O(4)$-invariant negative mode is sufficient \cite{Witten:1981gj,Coleman:1980aw}.  Here, the $O(4)$ symmetry of our instantons guarantees that fluctuations can be divided up by their transformation properties under $O(4)$, making $\mathcal{D}_E$ block-diagonal in representations of  this group.  Since fluctuations in $\rho$ transform trivially, to quadratic order they can only mix with other $O(4)$-invariant modes.  Metric fluctuations other than those in $\rho$ are $\sigma$-dependent away from the wall location. Hence it is reasonable that the corresponding gradient energy will force them have higher action than than fluctuations in $\rho$, causing them to have more positive eigenvalues of $\mathcal{D}_E$.  Thus we assume that $\rho$ is an eigenmode of $\mathcal{D}_E$, so the corresponding eigenvalue $\lambda_\rho$ is
\be
\lambda_\rho = \frac{\partial^2 S_E}{\partial \rho^2}
\ee
To show that a given solution represents a tunneling instanton, we must show that $\lambda_\rho$ is negative.  More detailed arguments which support this calculation are given in Appendix \ref{appendix}.

\begin{figure}
\begin{center}
\includegraphics[scale=1]{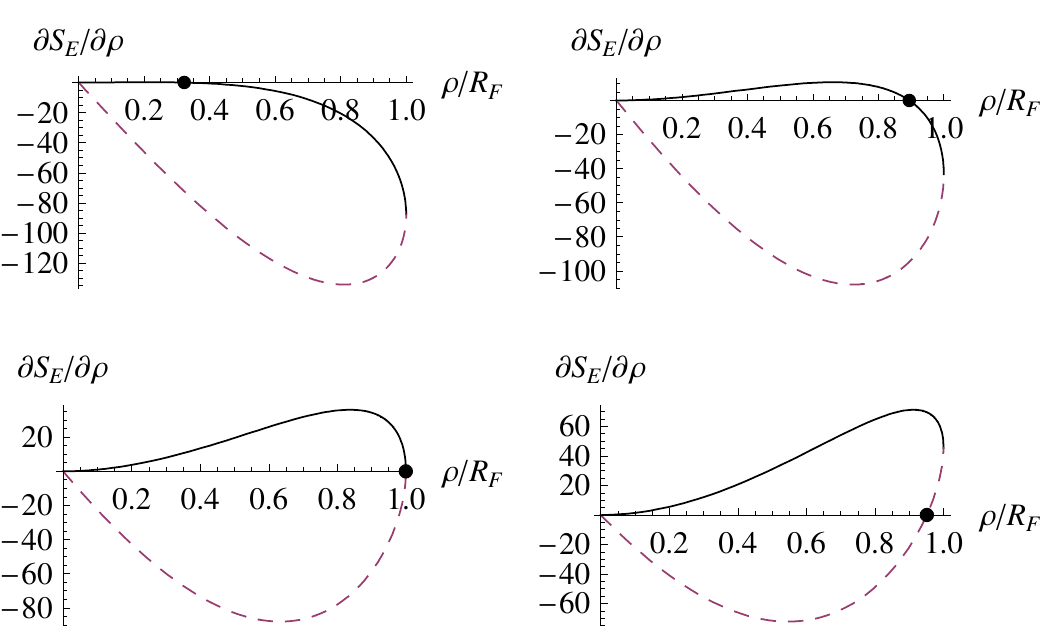}
\caption{First action derivative for positive tension instantons.  When the curve crosses the x-axis with negative slope, a tunneling instanton exists.  The solid line represents the derivative of the Type (a) instanton action with $(\mathcal{F},\mathcal{T})=(+1,-1)$, while the dashed line corresponds to Type (b), with $(\mathcal{F},\mathcal{T})=(-1,-1)$. Clockwise from top left corner: $T\ll T_{crit}$, $T<T_{crit}$,$T=T_{crit}$, and $T>T_{crit}$.  The tick numbering on the y-axis is arbitrary.}
\label{posTnegMode}
\end{center}
\end{figure}

To illustrate that this prescription gives the correct answer for the positive-tension case, the action derivatives are illustrated in Figure \ref{posTnegMode}  for varying values of the tension $T$.  When $T=0$ the join radius is zero, and as $T$ is increased to $T_{crit}$, the join radius $\rho$ is pushed to the false vacuum horizon $\rf$.  The stationary point $\rho$ minimizes the action and corresponds to the point at which the curves in Fig \ref{posTnegMode} cross the horizontal axis.  Below the critical tension, the slope at the stationary point is negative, indicating that a negative mode exists, and the corresponding solutions are tunneling solutions.  Above the critical tension, only actions of Type (b) have a stationary point, but the action derivative is increasing at these points, indicating the absence of a negative mode.  This analysis concludes that the Type (a) solutions (which include CdL instantons) represent tunneling solutions, and that Type (b) do not.

\begin{figure}[htbp]
\begin{center}
\includegraphics[scale=1]{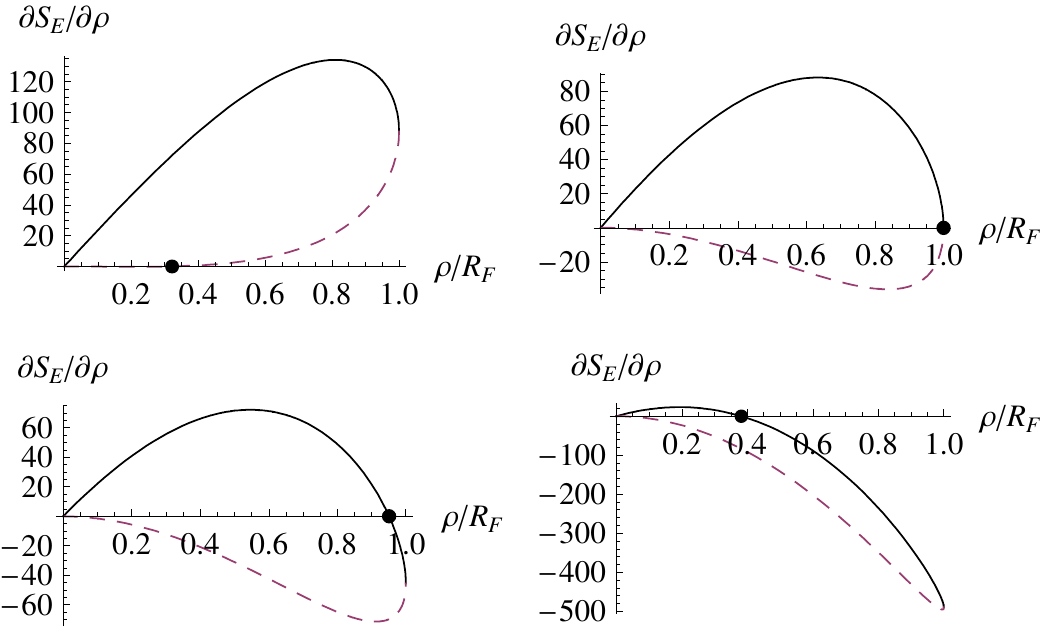}
\caption{First action derivative for negative tension instantons.  The dashed line represents the Type (c) instanton with $(\mathcal{F},\mathcal{T}) = (-1,+1)$, and the solid line represents the Type (d) instanton with $(\mathcal{F},\mathcal{T}) = (+1,+1)$.  Clockwise from top left corner: $T>-T_{crit}$, $T=-T_{crit}$,$T<-T_{crit}$, and $T<<-T_{crit}$.  For negative tension instantons, negative modes only exist for Type (d), when $T < -T_{crit}$.}
\label{negTnegMode}
\end{center}
\end{figure}

Euclidean action derivatives for negative values of $T$ are illustrated in Figure \ref{negTnegMode}.  The overall pattern is similar to that of the positive tension case.  However, now the negative mode appears as the magnitude of $T$ \emph{exceeds} this critical value and the instanton is described by Type (d), with $(\mathcal{F},\mathcal{T})=(+1,+1)$.  For the positive tension case, the bubble wall tension had to be less (in magnitude) than the critical tension in order for it to represent a tunneling solution.  For negative tension solutions, the bubble wall tension must be larger (in magnitude) than the critical tension.

The results above can be summarized by studying the explicit expression for the eigenvalue
\be\label{e:NegModeEx}
\frac{\partial^2 S_E}{\partial \rho^2} \bigg{|}_{\rho=\rho_{\rm nuc}} = - 12\pi^2 \left( \mathcal{F} \left[1-\left(\frac{\rho}{R_F}\right)^2 \right]^{-1/2} + \mathcal{T} \left[1-\left(\frac{\rho}{R_T}\right)^2 \right]^{-1/2}\right)
\ee
Since $\rho \le R_F \le R_T$, the first term in parentheses is always larger in magnitude than the second.  Therefore the sign of $\mathcal{F}$ completely controls the existence of the negative mode.  When $\mathcal{F} = -1$, there is no negative mode and the instanton does not represent a tunneling process.  When $\mathcal{F}=+1$, there is a negative mode, and the solution represents a valid tunneling process.   The quenching of the negative mode is closely related to the cosmological horizon structure, as we discuss more fully in Section \ref{ss:CausalStructure}.

The expression (\ref{e:NegModeEx}) is singular as the join radius $\rho$ approaches the false vacuum horizon $R_F$. This divergence is an artifact of the coordinate singularity in $\rho$ as the join surface approaches the equator of the false vacuum $S^4$.  A better-behaved coordinate is the proper radius $\sigma_{\rm nuc}$ of the bubble wall as from the center of the Euclidean false vacuum spacetime (the ``north pole" in Figure \ref{fourclasses}).  We can focus on nearly-critical bubbles by taking
\be
\sigma_{\rm nuc} = \left(\frac{\pi}{2}+\delta\theta\right)R_F
\ee
so that $\delta\theta = 0$ corresponds to $\rho = R_F$.  Then
\be
\frac{\partial^2 S_E}{\partial \sigma^2} \bigg{|}_{\sigma=\sigma_{\rm nuc}}
= -12 \pi^2 \,\mathcal{F}\, \delta\theta + \mathcal{O}(\delta\theta^2)
\ee
Again, the existence of the negative mode is controlled by $\mathcal{F}$. With the better-behaved coordinate $\sigma$, the eigenvalue goes to zero smoothly  as the bubble radius approaches $R_F$.

To summarize, the only instantons with negative modes, and hence the only tunneling instantons, are those of Type (a) (for positive tension) and Type (d) (for negative tension).  In order to have a negative mode, a positive tension instanton must have a sufficiently small tension: $T < T_{crit}$.  The new negative tension instantons have the curious feature that they require a tension that is sufficiently large in magnitude: $|T| > T_{crit}$ in order for the negative mode to exist.  Despite their strange properties, the negative tension instantons possess the same features -- they are finite-action solutions of the Euclidean equations of motion with negative modes -- as thin-wall CdL instantons.  Therefore they represent valid tunneling solutions, and in the next two sections, we give two constructions of this new family of tunneling instantons.

\section{Construction I: Ghost condensate}\label{s:ghost}

In this section we describe a realization of the negative tension domain wall by a scalar field theory with non-canonical kinetic term, and show how it can be used to build negative-tension instantons between Minkowski vacua.  This is a closer analogy with the classic Coleman-de Luccia bubble, where the wall tension in built from scalar field kinetic and potential energy, in a manner similar to the construction of domain wall solutions.  For canonical scalar fields, the bubble tension must always be positive.  Here, we show that it is possible to construct domain walls with negative tension when higher-derivative terms are present in the scalar field action.  We use the machinery of ghost condensates, which are reviewed in Section \ref{ss:GhostBasics}.  In Section \ref{ss:NTDomain} we give a detailed study of the conditions that the scalar kinetic term and potential must satisfy, and show how negative-tension domain walls may be constructed. We give an worked example of such a wall in Section \ref{ss:ExplicitNTD}, numerically integrate the relevant ODEs, and explain how this wall fits into the negative-tension instanton solution.  We also discuss the relation between the ghost condensate theory studied here, and those that have cosmological applications.

\subsection{The basics of ghost condensates}\label{ss:GhostBasics}

For our construction we draw heavily on the literature for ghost condensates, which are a family of higher-derivative scalar field actions that have NEC-violating solutions \cite{ArkaniHamed:2003uy}-\cite{ArmendarizPicon:2000dh}.   These theories have been extensively studied, especially in the homogeneous but time-dependent case relevant for cosmology, as well as in black hole solutions \cite{Mukohyama:2005rw}.  They form a keystone of some early-universe cosmological models, such as the new ekpyrotic scenario \cite{Buchbinder:2007ad} and others \cite{Creminelli:2006xe,Creminelli:2007aq}.  These theories may have problems with gravitational thermodynamics \cite{Rubakov:2004eb}-\cite{Eling:2007qd}, but here we merely employ them as a perturbatively stable example of higher-derivative models with interesting properties.

Ghost condensate actions have the general form
\be\label{e:GCaction}
S = \int \left[ P(X) - V(\vp) \right] \, \text{d}^4 x
\ee
where
\be\label{e:DefOfX}
X = -\frac{1}{2} \eta^{\mu\nu} \partial_\mu \vp \partial_\nu \vp
\ee
and we have used a flat metric $\eta^{\mu\nu}$ since in this section we confine our discussion to Minkowski space.  The stress-energy tensor for the ghost condensate is
\be\label{e:GhostT}
T_{\mu\nu} = \eta_{\mu\nu} P(X) - \eta_{\mu\nu} V(\vp)+ P_X(X) \partial_\mu \vp \partial_\nu \vp 
\ee
where $P_X = \text{d}P/\text{d}X$ and $P_{XX} = \text{d}^2P/\text{d}X^2$.  This stress-energy tensor can allow NEC violation.  If $n^\mu$ is any null vector, then
\be
n^\mu n^\nu T_{\mu\nu} = P_X ( n^\mu \partial_\mu \vp )^2
\ee
so any field configuration with $P_X < 0$ automatically violates the NEC.\footnote{The potential $V(\vp)$ is irrelevant for violating the NEC because it acts like a cosmological constant, which saturates the NEC inequality.}  To find a negative-tension domain wall we need a solution with $P_X < 0$.  If take a field configuration $\vp = \vp_0 + \pi$ expanded about a background solution $\vp_0$, then the fluctuations $\pi$ will have non-ghostly (proper sign) kinetic terms only if
\be\label{e:NoGhost}
2XP_{XX} + P_X > 0
\ee
We must always be careful to enforce this condition, since the presence of ghosts is a sure sign that the background is pathological.

\subsection{Construction of a negative-tension domain wall}\label{ss:NTDomain}

We begin by giving a domain wall construction from a scalar field with canonical kinetic term, which we then generalize to the ghost condensate action.  The scalar field action is
\be\label{e:RegScalarAction}
S = \int -\frac{1}{2}(\partial \vp)^2 - V(\vp) \; \text{d}^4 x
\ee
Since we are constructing a one-dimensional domain wall, we take $\vp$ to depend on one coordinate $z$ only, so $\vp = \vp(z)$.  The quantity
\be
\epsilon = -\frac{1}{2}(\partial_z \vp)^2 + V(\vp)   
\ee
satisfies
\be
\partial_z \epsilon = 0
\ee
as can be verified by the equations of motion following from (\ref{e:RegScalarAction}).  The quantity $\epsilon$ can also be constructed directly by Legendre transformation of the Lagrangian in (\ref{e:RegScalarAction}).  

Since $\epsilon$ is conserved, by fixing its value at infinity we obtain a solution for $\vp$ by quadrature.  We seek an asmptotically trivial solution, with vanishing energy density and pressure.  This implies $\epsilon$ must vanish, which follows from the stress-energy tensor corresponding to (\ref{e:RegScalarAction}), which is
\begin{subequations}\label{e:RegScalarT}
\begin{align}
T_{00} &= \frac{1}{2}(\partial_z \vp)^2 + V(\vp) \label{e:RegScalarT:00} \\
T_{zz} &= \frac{1}{2}(\partial_z \vp)^2 - V(\vp) = -\epsilon \label{e:RegScalarT:zz} \\
T_{xx} &= T_{yy} = -T_{00}
\end{align}
\end{subequations}
In order to have vanishing pressure at infinity $T_{zz}$ must be zero, and so from (\ref{e:RegScalarT:zz}) we see that $\epsilon$ must be zero.  Combining this with the vanishing of $T_{00}$, we derive the stronger requirement that both $V(\vp)$ and $\partial_z \vp$ must vanish individually.   Denoting the value of $\vp$ at infinity by $\vp_\infty$, $V(\vp)$ must be tuned so that it has a minimum at $\vp_\infty$ satisfying $V(\vp_\infty)=0$.  We must enforce two tunings: one of the value of the potential, and the other of the scalar field gradient at infinity. Otherwise there is no domain wall solution satisfying our requirement of vanishing stress-energy at infinity.

From (\ref{e:RegScalarT}), the configuration that results has the stress-energy of a codimension-one object with tension that is ``smeared" over the $z$-direction: the pressure perpendicular to the object vanishes, and the transverse pressures are all equal and of opposite sign to the energy density.  Furthermore, the domain wall always has positive tension.  This is because $\epsilon = 0$ implies that $V(\vp)$ cannot be negative. Then (\ref{e:RegScalarT:00}) implies that $T_{00}$, and thus the tension, is positive.  These walls exist for any choice of $V(\vp)$, provided it has at least two\footnote{Multiple minima are required for (meta)stability of the walls.} points with $V=0$, and is everywhere non-negative. Therefore we can only construct domain walls with positive tension from canonical scalar fields with potential.

We now show that negative tension domain walls can be constructed if the scalar field action has higher-derivative terms.  Our construction is a generalization of the one employed above for a canonical scalar field.  The new construction is more complex because additional complications arise when higher-derivative terms are present.  One issue is that higher-derivative actions can give rise to both positive and negative-tension walls, and we must ensure that the wall we construct has negative tension.  Another subtlety is the existence of pathological backgrounds with ghostly excitations, which must be avoided.  There are also cases in which the underlying fields can develop cusps, become multiply valued, or discontinuous.  Below we show how to construct a background containing a negative-tension domain wall in which all of these pathologies are absent.

We mimic the canonical scalar field case by constructing a conserved quantity.  The canonical momentum density $\Pi^\mu$ associated with $\vp$ is  
\be\label{e:GhostPi}
\Pi^\mu = \frac{\delta L}{\delta \partial_\mu \vp} = - P_X \eta^{\mu\nu} \partial_\nu \vp
\ee
The equation of motion is
\be\label{e:GhostEOM}
\partial_\mu \Pi^\mu + \frac{\partial V}{\partial \vp} = 0.
\ee
We assume that the solution is static ($\partial_0 \vp = 0$) and only depends on the $z$-coordinate.  We perform the Legendre transformation
\be\label{e:DefOfEps}
\epsilon = \Pi^z \partial_z \vp  - L = 2XP_X - P + V = -K + V
\ee
where the second equality follows from the assumption of stasis, and in the third we have defined 
\be
K= P - 2XP_X.
\ee 
The quantity $\epsilon$ is independent of $z$, as can be readily verified from the equations of motion (\ref{e:GhostEOM}).  It is the generalisation of the conserved $\epsilon$ introduced above for the canonical scalar field.

Since $\epsilon$ is independent of $z$, by analogy to the canonical scalar field, we should be able to solve for $\vp$ by quadrature.  This is possible provided $K[\partial_z \vp]$ is an invertible function of $\partial_z \vp$.  The inverse function theorem guarantees $K$ is invertible so long as we always remain in a region where the derivative of $K$ is nonvanishing.\footnote{We could have an inverse of $K$ if only its first derivative vanishes, but the inverse function ($\partial_x\vp$ in this case) would have infinite first derivative, which we exclude as pathological.}  Since
\be\label{e:SingleValued}
\frac{\text{d} K}{\text{d} \partial_z \vp} = 
\partial_z \vp \left( P_X + 2XP_{XX} \right)
\ee
we are safe provided $\partial_z \vp \ne 0$ and we remain within a ghost-free region where $P_X + 2XP_{XX} > 0$.

Thus far everything has been completely general, but for the sake of constructing an explicit solution, we now focus on a specific choice of $P(X)$, and show how the various constraints can be simultaneously satisfied.  This explicit choice serves as an existence proof, but we expect many other choices of $P(X)$ to work equally well. We suppose
\be\label{e:PX}
P(X) = -X + \frac{\alpha}{3} X^3
\ee
with $\alpha > 0$ a free parameter.  This $P(X)$ has extrema when $X$ is both spacelike ($X < 0$) and timelike ($X > 0$).  We focus on the spacelike extremum here, and will comment below on the relevance of the timelike one to cosmology.

Our choice of $P(X)$ only leads to healthy NEC-violating theories when certain consistency conditions, described above, are met.  First, we must be in a regime where there are no ghostlike excitations.  Using (\ref{e:NoGhost}) this requires
\be\label{e:PXNoGhost}
|\partial_z\vp| > \left( \frac{4}{5\alpha} \right)^{1/4}
\ee
For this model, 
\be\label{e:PXK}
K = -\frac{1}{2} \left(\partial_z \vp\right)^2
+ \frac{5\alpha}{24} \left(\partial_z \vp\right)^6
\ee
and (\ref{e:SingleValued}) implies that $K$ is guaranteed to be invertible when (\ref{e:PXNoGhost}) is satisfied.  Our choice of $P(X)$ violates NEC when $P_X < 0$ which corresponds to
\be\label{e:PXNoNEC}
|\partial_z\vp| < \left( \frac{4}{\alpha} \right)^{1/4} 
\ee
If this bound is violated, the NEC is satisfied.  Inspecting (\ref{e:PXNoGhost}) and (\ref{e:PXNoNEC}) it is clear that there is a ghost-free, but NEC-violating, choice for $\partial_z \vp$.

We now use boundary conditions at infinity to set the value of $\epsilon$.  We want the stress-energy tensor to vanish at infinity, and from (\ref{e:GhostT}) we have
\begin{subequations}\label{e:PXT}
\begin{align}
T_{00} &= -P + V \label{e:PXT00} \\
T_{zz} &= P - 2XP_X - V = K - V = -\epsilon \label{e:PXTzz} \\
T_{xx} &= T_{yy} = - T_{00} \label{e:PXTxx}
\end{align}
\end{subequations}
Since we want $T_{zz} = 0$ at infinity, $\epsilon = 0$ everywhere, as in the scalar field case.  Now by adding (\ref{e:PXT00}) and (\ref{e:PXTzz}) we find that $XP_X$ must vanish.  Since $X$ is nonzero by (\ref{e:PXNoGhost}), we must have $P_X = 0$ and so at infinity $X$ sits at an extremum of $P(X)$.  For our choice (\ref{e:PX}) this corresponds to 
\be\label{e:GradPhiInfty}
\partial_z \vp |_\infty = \left(\frac{4}{\alpha}\right)^{1/4}
\ee
This is on the boundary between the NEC-satisfying and NEC-violating regimes, but far from the ghostly regime.  Returning to the $\epsilon = 0$ condition, at this extremum $P(X)$ is nonzero, so we must also have
\be\label{e:VInfty}
V(\vp_\infty) = \frac{2}{3\sqrt{\alpha}}
\ee
Unlike the scalar field case, neither the gradient nor the potential vanishes at infinity.  In order for (\ref{e:GradPhiInfty}) and (\ref{e:VInfty}) to be mutually consistent, $V(\vp)$ must plateau at very large positive and negative values of $\vp$. The two tunings (\ref{e:GradPhiInfty}) and (\ref{e:VInfty}) are precisely the same number that we needed for the canonical scalar field. 

For almost any potential $V(\vp)$ satisfying (\ref{e:VInfty}) we can use the $\epsilon=0$ equation to find $\vp(z)$ by quadrature.  There is only one additional restriction.  For $\partial_z \vp$ in the range defined by (\ref{e:PXNoGhost}), $K$ must satisfy
\be
- \frac{2}{3\sqrt{5\alpha}} \leq K 
\ee
Therefore $V(\vp)$ must obey a similar bound
\be
 - \frac{2}{3\sqrt{5\alpha}} \leq V(\vp) 
 \ee
Except for this condition, $V(\vp)$ is arbitrary, provided its asymptotic behavior at $\vp = \pm \infty$ is as required by (\ref{e:VInfty}).  The restriction of the range of $V$ is precisely analogous to that for the canonical scalar field, where we required that $V$ be non-negative and asymptote to zero at $\varphi_\infty$.  

This family of ghost condensate solutions contains both positive and negative tension walls.  To ensure that the wall tension is negative definite, a sufficient condition is that $P_X < 0$ everywhere in the solution.  This implies the bounds
\be
\left(\frac{4}{5\alpha}\right)^{1/4} < |\partial_z \vp| < \left(\frac{4}{\alpha}\right)^{1/4}
\ee
which further imply
\be\label{e:PXonlyNEC}
- \frac{2}{3\sqrt{5\alpha}} < V < \frac{2}{3\sqrt{\alpha}}
\ee
for all $z$.
The condition (\ref{e:PXonlyNEC}) is sufficient but not necessary to obtain a negative tension wall. If $V$ exceeds the upper bound, the resulting solution will have both NEC-satisfying and NEC-violating phases.  These could balance out in different ways to give domain walls that have net positive or negative tension.  All this means is that our ghost condensate theory has both positive and negative tension domain wall solutions, and by applying (\ref{e:PXonlyNEC}) we can ensure that we only have negative-tension walls.

As the final step, we show that the stress-energy resulting from this construction is precisely that of a domain wall, but with negative tension.  Since $\epsilon=0$, (\ref{e:PXTzz}) shows that $T_{zz} = 0$ and so there is no pressure perpendicular to the wall.  Inspecting (\ref{e:PXT00}) and (\ref{e:PXTxx}) shows that the pressure transverse to the wall is equal but of opposite sign to the energy density.  This shows our configuration acts like an object with tension.  To show the tension is negative, we need to show that $T_{00} < 0$.  Since
\be
T_{00} = -P + V = -K-2XP_X + V = \epsilon - 2XP_X = -2XP_X
\ee
and $P_X < 0$ by assumption, and $X < 0$ because we are looking at spacelike field gradients, the right hand side is negative definite.  This means $T_{00} < 0$ and we have a negative-tension wall as desired.

\subsection{An explicit example}\label{ss:ExplicitNTD}

\begin{figure}
\begin{center}
\includegraphics{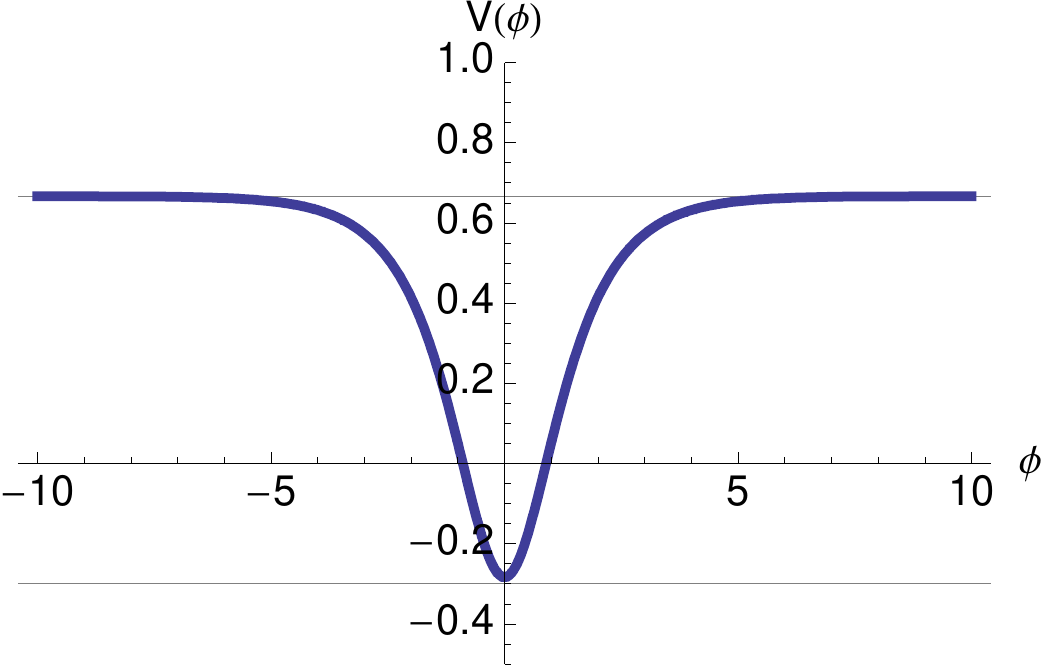}
\caption{The potential we use for explicitly constructing a negative-tension domain wall.  We set $\alpha=1$, $A = 0.95$, and $\Delta\varphi = 1$.  The top horizontal lines is at $V = 2/3$, the asmyptote for large and small $\vp$.  The lower horizontal line is at $V=-2/3\sqrt{5}$.  If the potential went below this line, the resulting wall tension would not be negative definite.}
\end{center}
\label{f:VOfPhi}
\end{figure}

\begin{figure}
\begin{center}
\includegraphics{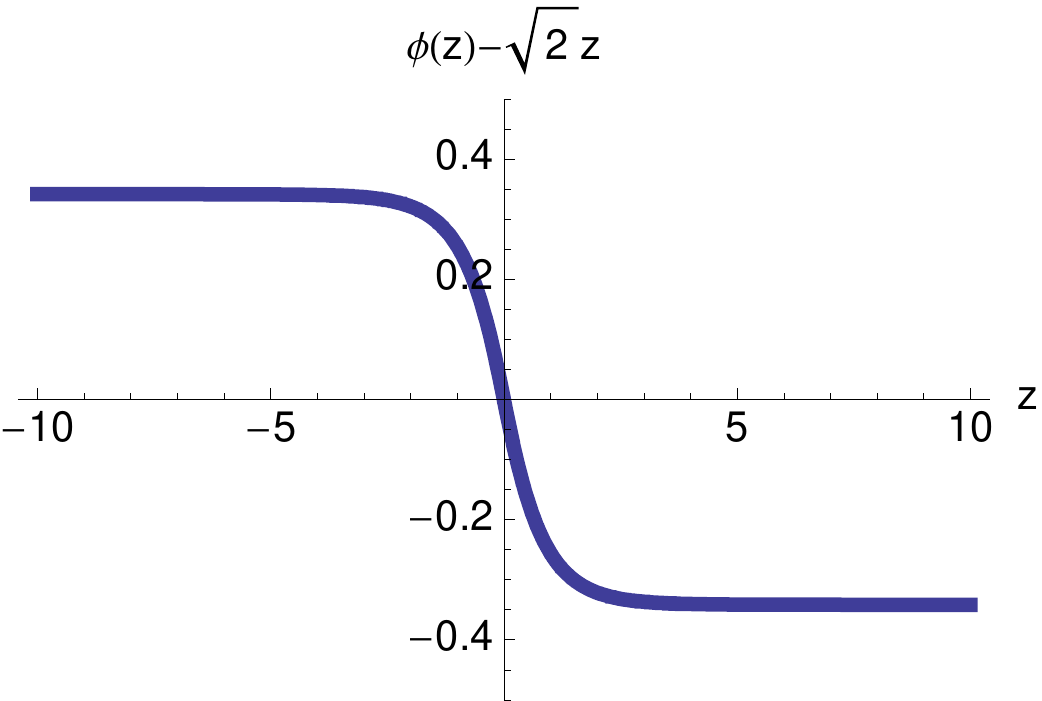}
\end{center}
\caption{Domain wall solution for $\vp(z)$ with $\alpha=1$, $A = 0.95$, and $\Delta\varphi = 1$ resulting from the potential shown in Figure 6.  For clarity, the plot shows the solution $\vp(z)$ with the linear term $\sqrt{2} z$ subtracted off.  The initial condition $\vp = -13.8005$ at $z = -10$ was set to center the feature near $z=0$.  Other choices for the initial condition would shift the location of the feature. }
\label{f:PhiOfz}
\end{figure}

Thus far we have kept the potential completely general modulo some inequality constraints.  We can show an explicit solution for a particular form of the potential which satisfies all of these constraints.  One example is
\be\label{e:exampleV}
V(\vp) = \frac{2}{3\sqrt{\alpha}} - \frac{A}{\cosh (\vp/\Delta\vp)}
\ee
This potential is not motivated by any more fundamental physics, but provides a simple example of a potential with  localized features and the required plateaus as large and small values of $\vp$.  The potential is illustrated in Figure \ref{f:VOfPhi}, and satisfies all of our constraints provided
\be
0 < A < \frac{2(5+\sqrt{5})}{15\sqrt{\alpha}} \simeq \frac{0.9648}{\sqrt{\alpha}}
\ee
Starting from (\ref{e:DefOfEps}), and using $\epsilon = 0$, we find $K=V$ everywhere.  Using this relationship, the expression (\ref{e:PXK}) can be inverted to give $\partial_z \vp$ in terms of $V$.  (The inversion can be done in closed form because (\ref{e:PXK}) is a cubic in $(\partial_z\varphi)^2$, but the explicit expression is rather complicated.)  This gives a first-order differential equation for $\vp$.  At very large positive or negative values of $\vp$, away from the domain wall, this differential equation implies
\be
\partial_z \vp(z) = \frac{\sqrt{2}}{\alpha^{1/4}} \qquad \text{(far from domain wall)}
\ee
so the solution is linear in $z$, with a feature near the location of the domain wall.  One such solution, obtained by numerical integration, is shown in Figure \ref{f:PhiOfz}.

\begin{figure}
\begin{center}
\includegraphics{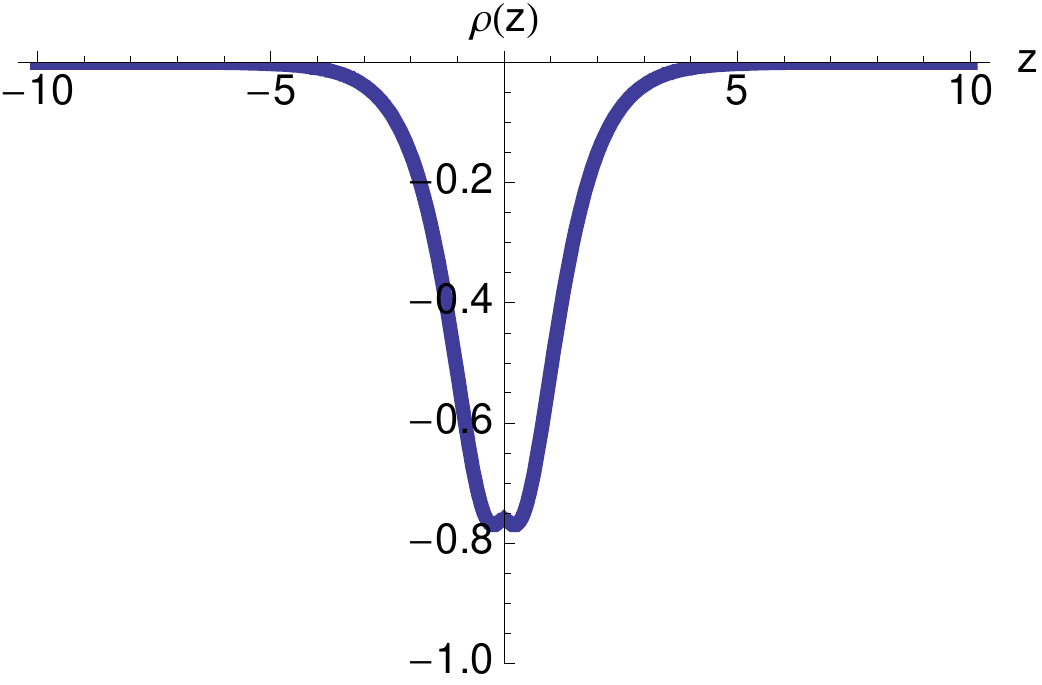}
\end{center}
\caption{Energy density $\rho$ for the domain wall solution for $\vp(z)$ with $\alpha=1$, $A = 0.95$, and $\Delta\varphi = 1$ using the numerical solution gained from the potential in Figure 6 and shown in Figure 7.  This energy density profile is precisely that of a domain wall but with negative tension.  Such a solution is impossible with a canonical scalar field, but here we see it can be constructed with higher-derivative kinetic terms.}
\label{f:RhoOfz}
\end{figure}

At this point the potential and scalar field profile are qualitatively similar to those for a domain wall built from a canonical scalar field.  The difference comes when we take this solution and use the expression for the energy density (\ref{e:PXT00}).  This gives a negative definite energy density profile as shown in Figure \ref{f:RhoOfz}.

In combination with the full expression (\ref{e:PXT}) for the stress-energy, this demonstrates that the numerical solution represents a domain wall with negative tension.  Such domain walls are impossible with a canonical scalar field, but are possible with the higher-derivative scalar field theories shown here.  The width of the domain wall $\Delta z$ can be estimated by dividing the width   $\Delta\phi$ of the ``bump" in the potential by the scalar field gradient $\partial_z\varphi$ at infinity, giving
\be\label{e:NTDwidth}
\Delta z \sim \frac{\Delta\varphi}{\partial_z\varphi |_\infty} \sim
\alpha^{1/4} \Delta\varphi
\ee
Therefore the negative-tension domain wall can be made as thin as desired by adjusting features in the kinetic term and potential.

We have constructed the required negative-tension domain wall, but the final step is to show that this can be ``lifted" to a full CdL-like bubble wall. First we establish the behavior of the $P(X)$ ghost condensate term under the analytic continuation $t \to i\tau$. In the $O(4)$ symmetric situation in which $\vp$ is a function of the Euclidean radius $\rho$, given by
\be
\rho = \left( \tau^2 + \vec{x}^2 \right)^{1/2}
\ee
analytic continuation gives
\be
X = -\frac{1}{2}\left( (\partial_\tau \vp)^2 + (\nabla \vp)^2 \right) 
= - \frac{1}{2} \left( \frac{\td \vp}{\td \rho} \right)^2
\ee
Therefore $X$ has the same sign as it did in our domain wall solutions, and in fact the $P(X)$ term only changes by the formal replacement of $z$ by the Euclidean radius $\rho$.  The only other change from our domain wall example is that the action (\ref{e:GCaction}) becomes
\be
S_E =  \Omega_{(3)} \int  \rho^3 \left[ P(X) - V(\vp) \right] \, \text{d} \tau 
\ee
where $\Omega_{(3)}$ is the area of the unit $S^3$.  The new volume measure modifies the equations of motion (\ref{e:GhostEOM}) to
\be\label{e:GhostEOMdamping}
\partial_\mu \Pi^\mu + \frac{3}{\rho}\Pi^\rho + \frac{\partial V}{\partial \vp} = 0
\ee
where we have kept the old definition (\ref{e:GhostPi}) of $\Pi^\mu$.  Just as in the traditional CdL instanton, the second term in (\ref{e:GhostEOMdamping}) is negligible in the thin-wall limit where the wall thickness is much less than the bubble $\rho$ at nucleation.  Using the estimate (\ref{e:NTDwidth}) the width of the domain wall can be made as small as desired by an appropriate choice of parameters in the kinetic term and potential, so it is always possible to remain in the thin-wall limit.  The domain wall construction assumed that space was asymptotically Minkowski, but if space is asymptotically de Sitter this cannot influence the domain wall solution so long as the de Sitter horizon radius is much larger than the domain wall thickness, as it will be in the thin-wall limit.  Furthermore, unlike positive-tension instantons, there exist negative-tension instantons that transition between two Minkowski vacua, in which case the domain wall solution is exact in the thin-wall limit (See Section \ref{ss:Z2Mink}).  We conclude that, by choosing parameters so that $\Delta z$ is sufficiently small, our negative-tension domain wall solution gives a thin-wall negative-tension tunneling solutions if we replace $z \to \rho$.

While it was chosen to illustrate the existence of negative tension domain walls, the ghost condensate (\ref{e:PX}) has interesting properties from a cosmological perspective.  One application of ghost condensate theories is in constructing nontrivial field theories with give de Sitter minima.  For these solutions the background solution is a scalar field which is evolving linearly with time.  Our $P(X)$ has such an extremum for positive $X$, at $X =  \alpha^{-1/2}$.  This extremum is ghost-free.  Furthermore, it produces a positive (de Sitter) cosmological constant.  Using the same potential as we have employed for the domain wall, and assuming we are at a background value of $\vp$ on the plateau, at this other minimum we have
\be
\rho = \frac{4}{3\sqrt{\alpha}}, \quad P_x = P_y = P_z = -\rho
\ee
corresponding to a de Sitter phase.  The trajectory we have studied for the domain wall is the spacelike analogue of a brief NEC-violating phase which begins, and ends, in a de Sitter vacuum.

We have argued that theories which have negative tension domain walls can be unstable to bubble nucleation.  In this section we have shown that it is possible to construct such domain walls in some higher-derivative scalar field theories, and in particular theories which are superficially similar to those used to engineer NEC-violating phases in some cosmological models.  What is less clear is whether the requirement of a cosmological NEC-violating solution (which depends on $P(X)$ for $X>0$) necessarily implies there are NEC-violating domain walls (which depend on $P(X)$ for $X<0$.)  In this example we have seen that both cosmological and domain wall NEC-violating phases are present in the same model.  Unfortunately, these two phases seem to be separated by a ghostly background: a point at which the scalar field configuration becomes part of a domain wall, and so switches from $X>0$ to $X<0$, must pass through the $X=0$ which has ghost excitations.  This probably means that this trajectory should be forbidden classically: however, quantum-mechanically it might be possible to transition from a de Sitter phase to a domain wall phase.  This would correspond to a complete analogue of CdL de Sitter $\to$ de Sitter tunneling with negative tension bubble.  Having demonstrated a Minkowski $\to$ Minkowski tunneling solution here, we leave the construction of a full de Sitter $\to$ de Sitter tunneling solution to future work.

\section{Construction II: Negative-tension $Z_2$ boundary}\label{s:orientifold}

Our second construction of negative-tension instantons uses some exotic objects, similar to those that form crucial parts of string and supergravity constructions, as well as some well-known higher-dimensional models.  We take as our model theories in which a spacetime boundary arises as a $Z_2$ orient- or orbifold of a boundary-free space, with the resulting boundary carrying negative tension, as reviewed briefly in Section \ref{ss:NTB}.  The resulting instanton describes the formation of a ``bubble of nothing" which expands at nearly the speed of light and consumes a horizon volume.  We construct the solutions for de Sitter and Minkowski space in Sections \ref{ss:Z2dS} and \ref{ss:Z2Mink} and show that the bubble nucleation rate is small.  Our solutions are similar to a ``bubble of nothing" found by Witten in the context of five-dimensional Kaluza-Klein theory, and we discuss the relation between the two in Section \ref{ss:WBON}.

\subsection{Negative-tension boundaries}\label{ss:NTB}

Some interesting higher-dimensional models include a boundary of spacetime, which often arises from the quotient of a boundary-free space by a discrete group with fixed points (orbi- or orientifolding).  In some cases, global consistency conditions require that there be stress-energy localized on these boundaries, which may correspond to positive or negative tension objects.  This class of theories have actions of the schematic form
\be\label{e:ActionWithBoundaries}
S_{\rm tot} = \frac{1}{2}
\int_\mathcal{M} R \sqrt{-g} \, \td^n x + +\sum_{i} \int_{\mathcal{B}_i} K \sqrt{h_i} \, \td^{n-1} x + \sum_{i} T_i \int_{\mathcal{B}_i} \sqrt{h_i} \, \td^{n-1} x
+ S_{\rm matter}
\ee
where $\mathcal{M}$ is the spacetime manifold with metric $g$, and $i$ indexes the boundary components $\mathcal{B}_i$, each of which carries an induced metric $h_i$ and an effective tension $T_i$.  The matter action $S_{\rm matter}$ can include fields in the ``bulk" $\mathcal{M}$ or localized on the boundaries $\mathcal{B}_i$.  The key point is that some of the $T_i$ may be negative.

There are several interesting examples of negative-tension spacetime boundaries.   They are useful for producing warped compactifications, the canonical example being the Randall-Sundrum model \cite{Randall:1999ee,Randall:1999vf}.  The orientifold planes of string theory provide another example \cite{Polchinski}.  In the simplest cases orientifolds arise from modding out spacetime by a $Z_2$ reflection symmetry combined with a worldsheet parity transformation.  These objects have many mysterious properties -- for example there are no string states bound to them and so they cannot fluctuate, as can D-branes and other extended objects which appear in string theory.  Orientifolds typically have negative tension and are useful building blocks in string models, especially in warped flux compactifications \cite{Verlinde:1999fy}-\cite{Burgess:2006mn}.  Negative-tension boundaries also appear in supergravity constructions (for example \cite{Bergshoeff:2000zn,Lukas:1998yy}) and have interesting behavior near singularities \cite{Lehners:2007nb} which may be useful for cosmology \cite{Lehners:2007ac,Lehners:2006ir}.

In this section we consider theories with actions of the form (\ref{e:ActionWithBoundaries}) and use them to construct a CdL-like instanton.  In the absence of a string formulation, we have no equivalent of the worldsheet parity operator, so we will search for solutions which possess a $Z_2$ symmetry.  Such solutions can be regarded as the boundaries of spacetime, in analogy with the orientifold planes of string theory.  If a theory can be written in the form (\ref{e:ActionWithBoundaries}) and possesses static negative tension boundaries, then there is no \emph{local} restriction which can prevent bubble-like solutions.  A tiny patch of bubble wall is identical to a (possibly boosted) tiny patch of static spacetime boundary.  There could be \emph{global} constraints in theories of the type (\ref{e:ActionWithBoundaries}) which prevent such bubbles from forming.  One example is a conserved charge carried by the boundaries.  (This ingredient can ensure stability as in \cite{Lehners:2005su}).  Here we consider only the simplest possibility where there are no such constraints.  We show that, using these negative-tension boundaries, it is possible to construct instantons representing non-perturbative instabilities in the theories which support them.

\subsection{de Sitter bubble}\label{ss:Z2dS}

\begin{figure}
\begin{minipage}[t]{4cm}
\begin{center}
\includegraphics[width=4cm,clip]{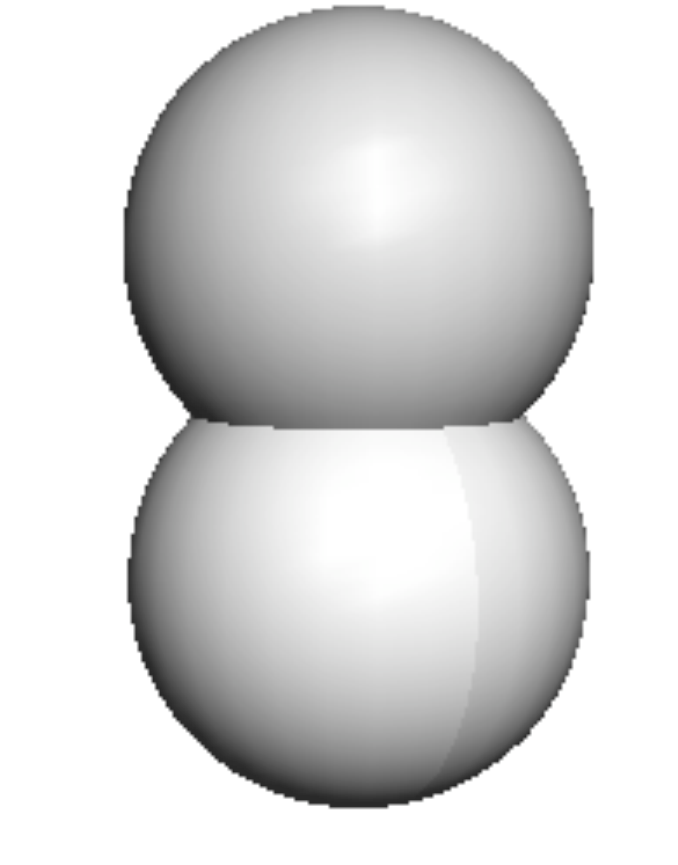}
\caption[Short caption for figure 1]{\label{nz2} Tunneling between degenerate vacua is enabled with a negative tension boundary}
\end{center}
\end{minipage}
\hfill
\begin{minipage}[t]{5cm}
\begin{center}
\includegraphics[width=4cm,clip]{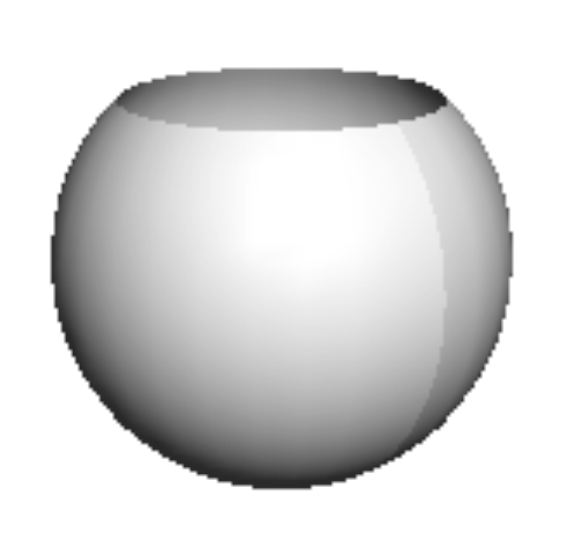}
\caption[Short caption for figure 2]{\label{orientifold}$Z_2$ identification and the spacetime boundary.}
\end{center}
\end{minipage}
\end{figure}

To ensure a $Z_2$ symmetry suitable for orbifolding, we require identical cosmological constants $\Lambda=\Lambda_F=\Lambda_T$ in each vacuum (so $\rf=\rt=R$ as well).  Because the cosmological constants are equal, the critical tension $T_{crit}$ vanishes.  The relevant tunneling instantons are of Type (a) for positive tension, and Type (d) for negative tension.

There are a number of reasons why the positive tension Type (a) tunneling solution should be discarded.  From (\ref{e:TensionRFRT}), this solution satisfies Israel matching only in the trivial case $T=0$.  Therefore the instanton is a Euclidean $S^4$ in every way identical to the ``false-vacuum" $S^4$, and this is simply the trivial, no-bubble solution.  For equal de Sitter radii, the critical tension (\ref{e:Tcrit}) vanishes, and since Type (a) has positive tension, the discussion in Section \ref{ss:NegModes} indicates that the required negative mode does not exist.  Therefore there are no thin-wall positive-tension tunneling instantons between de Sitter spaces with identical cosmological constants.

The situation is different for the negative-tension Type (d) tunneling solutions.  These correspond to different choices of $\mathcal{F}$ and $\mathcal{T}$ so the Israel matching condition (\ref{e:TensionRFRT}) is easily satisfied.  The negative mode discussion in Section \ref{ss:NegModes} indicates that the tension must be greater (in absolute value) than the critical tension: combined with the vanishing of the critical tension, this implies that a negative mode exists for any negative value of the tension $T$.  This means that for any choice of $T$ and $\Lambda$ there is always a tunneling instanton that satisfies the Euclidean equations of motion.  Therefore, in contrast to the positive-tension case, there is always a tunneling instanton between two degenerate vacua with negative tension.  

The Euclidean $Z_2$-symmetric negative-tension instanton is illustrated in Figure \ref{nz2}.  Since we are allowing only negative-tension boundaries that result from $Z_2$ orbifolding, we identify points related by the $Z_2$ reflection symmetry through the join.  This results in a space with boundary, illustrated in Figure \ref{orientifold}.  This boundary is locally the fixed point locus of a $Z_2$ symmetry and so arises in the same way as the negative-tension boundaries discussed in Section \ref{ss:NTB}.

To an observer in the de Sitter space with cosmological constant $\Lambda$, this instanton appears as the nucleation of a  closed spherical boundary.  The nucleation radius is obtained from (\ref{e:NucRadius}) and is
\be
\rho_{\rm nuc} = R \left[ 1 + \left(\frac{R T}{4 M_{pl}^2}\right)^2 \right]^{-1/2}
\ee
Once nucleated, there is nothing inside the bubble -- no spacetime and no matter fields.  It is a boundary of spacetime.  As in standard CdL tunneling, the equations of motion for the bubble wall imply that it expands at a speed rapidly approaching that of light, eventually filling an entire de Sitter horizon.  We discuss the Lorentzian continuation of this solution in more detail in Section \ref{s:interpretation}.  

We must be careful in computing the tunneling rate.  This is because we need to stay consistently in either the ``upstairs" (before the $Z_2$ quotient) or ``downstairs" (after the $Z_2$ quotient) picture.  If we decide to stay in the downstairs picture, the prescription (\ref{e:Bprescription}) is modified to
\be
\Gamma \sim \exp\left( -\frac{1}{2} S_E + S_{stay} \right)
\ee
where $S_E$ is the Euclidean action for the full ``upstairs" instanton configuration shown in Figure \ref{nz2}.  This gives
\be\label{e:GammaZ2dS}
\Gamma \sim \exp\left( - \frac{12\pi^2 M_{pl}^2}{\Lambda}\left[ 1 - \left( 1 + \frac{16 M_{pl}^4 \Lambda}{3T^2}\right)^{-1/2} \right] \right)
\ee
The argument of the exponential is negative definite.  This means that the tunneling rate is suppressed.  Even though we are dealing with negative-tension instantons, the $Z_2$ symmetry seems to eliminate the exponential enhancement that we might otherwise expect.

\subsection{Minkowski bubble}\label{ss:Z2Mink}

In this section, we construct the instanton which is the quotient of the negative-tension solution in which both the ``true" and ``false" vacua are Minkowski space.  Tunneling from de Sitter to Minkowski was studied in the original work on positive-tension instantons, but the Israel matching conditions forbid  positive-tension instanton solutions which tunnel between Minkowski vauca.  The existence of solutions which tunnel between Minkowski vacua is a unique feature of the negative-tension instanton solutions.  

\begin{figure}
\begin{center}
\includegraphics{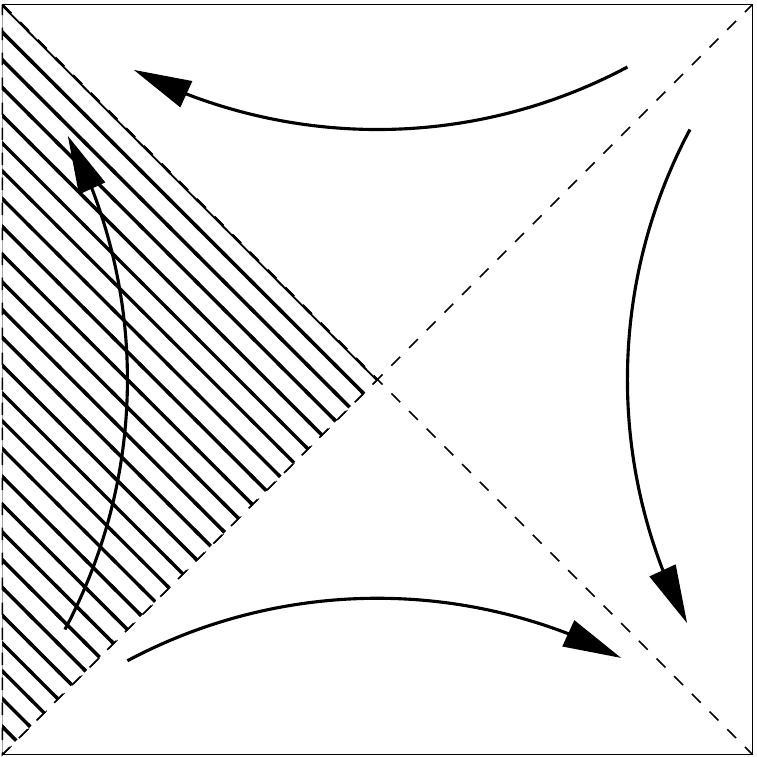}
\caption{Southern causal diamond in static coordinates (shaded region).  The thick lines indicate the flow of the timelike Killing vector in each quadrant.}
\label{staticcoords}
\end{center}
\end{figure}

To study the negative-tension boundary instanton in Minkowski space, we should be wary of simply taking the $\Lambda \to 0$ limit of (\ref{e:GammaZ2dS}).  The result  (\ref{e:GammaZ2dS}) appears well-defined in this limit, but the spacetime metric in global coordinates (\ref{global}) blows up.  This reflects the topology change required to flatten a sphere into a plane.  Additionally, taking $\Lambda \to 0$ in the global coordinate system (\ref{global}) leaves us with a spacetime whose conformal structure is different from Minkowski space.  For de Sitter space, null and timelike future infinities coincide on the spacelike surface $\mathcal{I}_+$, while in flat space null infinity is a null line and timelike infinity the point $i_0$.  To make the limiting process rigorous, we adopt the static slicing of de Sitter
\be
\td s^2 = - \left( 1-\frac{r^2}{R^2} \right) \td t^2 +  \left( 1-\frac{r^2}{R^2} \right)^{-1} \td r^2+ r^2(\td\theta^2 + \sin^2 \theta \, \td\phi^2).
\ee
These coordinates do not cover the entire manifold, but describe only the causal diamond in which the vector $\frac{\partial}{\partial t}$ is Killing (Figure \ref{staticcoords}). 
They are inadequate to describe de Sitter tunneling because the manifold extends beyond $r=R$, but the bubble wall reaches this coordinate boundary in finite time. However this coordinate system does have a well-defined Minkowski limit.    In the limit $R \rightarrow \infty$ the coordinate boundary coincides with the future null infinity of Minkowski space.

We can estimate the tunneling rate for these instantons by using the prescription
\be
\Gamma \sim \exp \left( - \frac{1}{2} S_E + S_{stay} \right)
\ee
but now only working in the causal patch defined by the static de Sitter coordinates.  This calculation has a well-defined $\Lambda \to 0$ limit.   

 We begin by rotating the static time coordinate $t \rightarrow i \xi$ to give a Euclidean manifold.   The action is once again described by the Einstein-Hilbert action but now we must be careful to include the Gibbons-Hawking term 
 \be
S_E \sim -\frac{\mpl}{2} \int \tilde{R} \sqrt{-g} d^4 x  -\mpl \int_{wall} K \sqrt{h} d^3 x+T\int_{wall} \sqrt{h} d^3 x
 \ee
 to properly account for the curvature near the bubble wall \footnote{ We accomplished this in Eq. (\ref{prettyaction}) by integration by parts, effectively adding a term of the form $K= 3 \frac{b'}{b}$.}.  In the flat-flat limit, the Ricci scalar $\tilde{R}$ is zero, so the only contribution to the first term comes from the extrinsic curvature of the boundary surface.  In the static coordinates it takes the form 
\be
\text{tr} \; K =3 \sqrt{\frac{1}{\rho^2} - \frac{1}{R^2}},
\ee
which reduces to $\text{tr} K = 3/\rho$ in the  flat-space limit.  Using this, the Euclidean action for the flat-space limit of the $Z_2$ instanton is
\be
\frac{S_E}{2\pi^2} = T \rho^3 + 6 M_{pl}^2 \rho^2.
\ee
Demanding this action be stationary gives the nucleation radius
\be
\rho = \frac{4 M_{pl}^2}{|T|}
\ee
and evaluating the Euclidean action then gives the tunneling rate
\be\label{e:GammaZ2Mink}
\Gamma \sim \exp\left( - \frac{32\pi^2}{T^2} \right).
\ee
The rate in fact agrees precisely with the $\Lambda\to 0$ limit of the de Sitter rate (\ref{e:GammaZ2dS}).  (When $\Lambda \ne 0$ the two rates disagree, reflecting the different causal structures of the instantons). As in the de Sitter case, the nucleation rate is exponentially suppressed, like the traditional CdL positive-tension instantons.

\subsection{Relation to Witten's ``bubble of nothing"}\label{ss:WBON}

Strange as our $Z_2$ negative-tension instantons are, they share some features with well-known solutions, such as the  ``bubble of nothing" described by Witten \cite{Witten:1981gj}.  Like the $Z_2$ symmetric solutions, the ``bubble of nothing" is a ``hole" in an asymptotically Minkowski spacetime which nucleates and grows.  Unlike our $Z_2$ solutions, in four dimensions the ``bubble of nothing" appears as a hole in spacetime surrounded by a particular scalar field profile.  Since canonical scalar fields cannot give negative-tension bubble walls, if this were a thin-wall instanton, it should not represent a valid tunneling solution, according to our analysis in Section \ref{s:review}.  In this section we resolve this puzzle.  We show that the ``bubble of nothing" has a positive energy density on its surface, but possesses a radial pressure, and so the bubble wall is not described uniquely by a surface tension.  This means the bubble of nothing does not fall under the cases studied in Section \ref{s:review}, and our conclusions are not in conflict with it representing a valid tunneling solution.

The Witten bubble is five-dimensional spacetime with metric
\be\label{e:WittenBubble}
\text{d}s^2 = 
\frac{\text{d}r^2}{(1-(R/r)^2)} - r^2 \,\text{d}\psi^2 + \cosh^2 \psi \,\text{d}
\Omega^2 + (1-(R/r)^2) \,\text{d}\phi^2
\ee
where $\text{d}\Omega^2$ is the metric on the unit two-sphere
\be
\text{d}\Omega^2 = \text{d}\theta^2 + \sin^2\theta \,\text{d}\chi^2
\ee
The metric (\ref{e:WittenBubble}) is a solution to the vacuum five-dimensional Einstein equations.  In order to avoid a conifold-type singularity, one must take $\phi$ to be a periodic variable with periodicity $2\pi R$.  Therefore, at large $r$, the solution (\ref{e:WittenBubble}) asymptotes to five-dimensional Kaluza-Klein spacetime, 
\be\label{e:KK}
\text{d}s^2 = 
\text{d}r^2 - r^2 \,\text{d}\psi^2 + \cosh^2 \psi \,\text{d}\Omega_3^2 + \text{d}\phi^2
\ee
with flat four-dimensional noncompat space and a circular extra dimension $\phi$ of radius $R$.   This solution has the curious feature that there is no spacetime for $r < R$, and so the solution describes a hole in spacetime.  As the radial coordinate $r$ approaches $R$, the radius of the fifth dimension $\phi$ shrinks to zero in just such a way so that the spacetime described by (\ref{e:WittenBubble}) is nonsingular and geodesically complete.

To a four-dimensional observer, this solution appears as a hole which has radius $R$ at $t=0$.  Inside the hole there is no spacetime -- the edge of the hole is a boundary for the four-dimensional universe.  The hole radius grows at a speed rapidly approaching that of light, and eventually consumes the entire universe.

This solution is interesting because it can represent the end state of a tunneling process: the flat Kaluza-Klein spacetime (\ref{e:KK}) can tunnel into the spacetime (\ref{e:WittenBubble}). The Euclidean action for fluctuations around (\ref{e:WittenBubble}) possesses a negative mode, and so it can represent a tunneling instanton, as shown in \cite{Witten:1981gj}.   Unlike the purely four-dimensional case, such a tunnelling process from Minkowski space is allowed by energy conservation.  The natural definition of energy in asmptotically flat 3+1 dimensional spacetimes, the ADM mass, is defined by the coefficient of $1/r$ terms in the 3+1 dimensional metric.  The ADM mass vanishes for both (\ref{e:KK}) and (\ref{e:WittenBubble}).  The existence of non Minkowskian solutions with vanishing ADM mass is closely related to the topology of the Kaluza-Klein solution (\ref{e:KK}) as discussed in detail in \cite{Witten:1981gj}.

These solutions are interesting precedents for the instantons we discuss in this paper.  We can compare the two solutions by going to the four-dimensional Einstein frame.  This is the conformal frame in which the four-dimensional effective $G_N$ is constant in both space and time, and is the natural frame for four-dimensional observers to use.  This metric in this frame is obtained by restricting the indices of (\ref{e:WittenBubble}) to the four noncompact dimensions $(r,\psi,\theta,\chi)$, and multiplying the restricted metric by the square root of the five-dimensional volume.  Up to an irrelevant positive constant, this gives
\be\label{e:WittenBubble4D}
\text{d}s^2_{4D} = 
\frac{\text{d}r^2}{(1-(R/r)^2)^{1/2}} + (1-(R/r)^2)^{1/2}\left( - r^2 \text{d}\psi^2 + \cosh^2 \psi \text{d}
\Omega^2 + (1-(R/r)^2) \text{d}\phi^2 \right)
\ee
To find the four-dimensional stress-energy corresponding to (\ref{e:WittenBubble4D}) we substitute this metric into the four-dimensional Einstein equations to obtain
\begin{subequations}
\begin{align}
-{G_\psi}^\psi &= \frac{3 R^4}{r^3(r^2 - R^2)^{3/2}} = \rho(r) \\
{G_r}^r &= P_r = \rho(r) \\
{G_\theta}^{\theta} &= P_\theta(r) = -\rho(r) \\
{G_\chi}^{\chi} &= P_\chi(r) = -\rho(r) 
\end{align}
\end{subequations}
which gives the energy density $\rho$ and pressures $P_r$, $P_\theta$, and $P_\chi$ along the $r,\theta$ and $\chi$ directions respectively.  This stress-energy is produced by the Kaluza-Klein scalar field profile around the bubble.  The stress-energy components are all $r$-dependent but stand in a fixed ratio with the energy density $\rho(r)$.  In terms of the physical distance $\ell$ from the bubble wall
\be
\ell = \sqrt{r^2 - R^2}
\ee
far from the wall, as $\ell \to \infty$, we have
\be
\rho(r) = \frac{3 R^4}{\ell^6} + \mathcal{O}(\ell^{-8})
\ee
while close to the wall, as $\ell \to 0$, we have
\be
\rho(r) = \frac{3 R}{\ell^3} + \mathcal{O}(\ell^{-1})
\ee
Observers in four dimensions see a bubble of nothing surrounded by a positive energy density which diverges rapidly (as $1/\ell^3$) close to the bubble wall, and decays quickly (as $1/\ell^6$) far from the wall.  There is a positive radial pressure perpendicular to the bubble wall, and negative pressure transverse to the wall.  This energy density and pressure is due to the gradient energy of the Kaluza-Klein scalar.

The stress-energy implies that, in four dimensions, the Witten bubble is akin to an bubble instanton with a positive tension wall.  It has pressures transverse to the wall of equal magnitude but opposite sign to the energy density, characteristic of objects with tension.  However it has positive radial pressure and so the bubble wall cannot be viewed purely as an object with tension.  That the Witten bubble does not appear as a purely positive-tension bubble wall solution is consistent with what we have shown above.  If the Witten bubble has a purely positive tension wall, our analysis would show that there would be no negative mode, and hence no instanton, for tunneling between two asmyptotically Minkowski vacua.  As it stands, none of these arguments conflict with the fact that the bubble of nothing appears to be a perfectly valid, if strange, tunneling solution.  

A similar analysis could be carried out for a M theory bubble of nothing described in  \cite{Fabinger:2000jd,Horava:2007hg,Horava:2007yh}.  In the lower-dimensional theory, this bubbles would appear as a hole in spacetime surrounded by positive energy density.  The solution of \cite{Fabinger:2000jd,Horava:2007hg,Horava:2007yh} involves a $Z_2$ projection, much like the solutions we describe in Sections \ref{ss:Z2dS} and \ref{ss:Z2Mink}.  However, in \cite{Fabinger:2000jd,Horava:2007hg,Horava:2007yh} the $Z_2$ acts along the compact extra dimension transverse to the radial direction from the bubble center, while in our solutions the $Z_2$ acts along the radial direction, with a fixed point locus on the bubble surface itself.  

\section{Causality protection}\label{s:interpretation}

We showed in Section \ref{ss:NegModes} that the existence of the $O(4)$-symmetric negative mode, and hence the interpretation of the Euclidean solution as a tunneling instanton, is controlled by the sign of $\mathcal{F}$.  In this section we show that this negative mode is deeply linked with causality, and ensures that the tunneling solutions are never acausal.  In Section \ref{ss:CausalStructure} we analytically continue the Euclidean solutions to Lorentzian signature to find their conformal structure.  We use this conformal structure in Section \ref{ss:CausalityNegModes} to show that the negative mode is only present for causal tunneling solutions.

\subsection{Lorentzian continuation and causal structure}\label{ss:CausalStructure}

\begin{figure}
\begin{center}
\includegraphics[scale=.4]{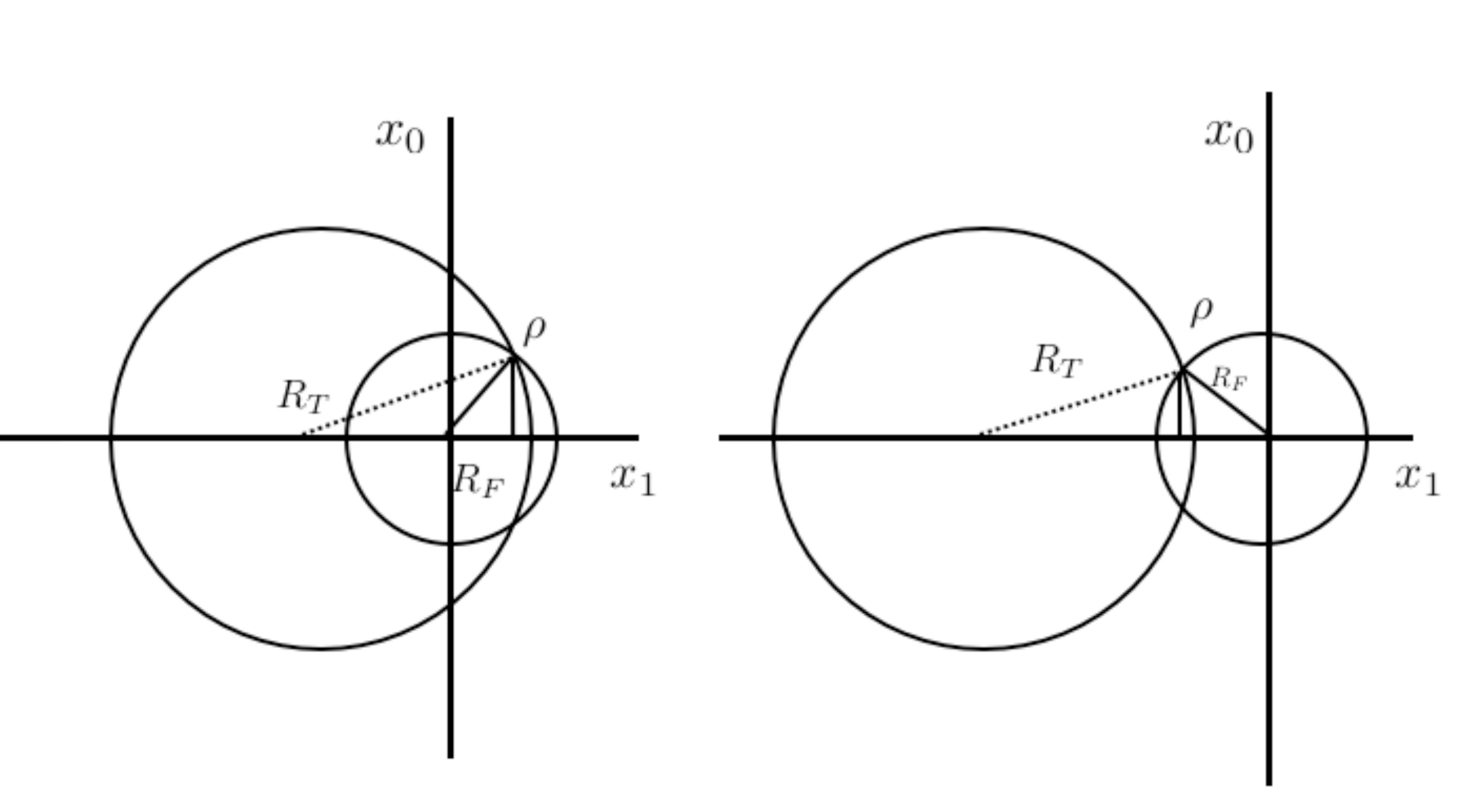}
\caption{The four instanton classes.  Left diagram: Type (a) corresponds to the inner surfaces, and Type (c) to the outer surfaces.  Right diagram: Type (b) corresponds to the inner surfaces, and Type (d) to the outer surfaces.  In either diagram, the positive tension instantons are the inner surfaces and the negative-tension instantons the outer.  The $x_1$ coordinate of the ambient Euclidean space is chosen to lie along the symmetry axis.}
\label{circles}
\end{center}
\end{figure}

We begin by analytically continuing each of the four classes of instanton to Lorentzian signature $(-+++)$.  The Euclidean false-vacuum $S^4$ is parameterized by the coordinate $\omega$ and three angles $\psi$, $\chi_1$, and $\chi_2$.  The pair $(\omega,\psi)$ are determined by the embedding of the $S^4$ in the ambient Euclidean space, with coordinates $x_0, \dots x_4$ centered on the false vacuum $S^4$ and given by
\begin{subequations}
\begin{align}
x_0 &= R_F \cos(\omega/R_F) \\
x_1 &= R_F \sin(\omega/R_F) \cos(\psi)\\
x_2 &= R_F \sin(\omega/R_F) \sin(\psi) \cos(\chi_1)
\end{align}
\end{subequations}
as illustrated in Figure \ref{circles}.
Henceforth we set $\chi_1 = 0$, and suppress the remaining Euclidean coordinates $x_3$,$x_4$ and the angles $\chi_1$,$\chi_2$ in the following.

\begin{figure}
\begin{minipage}[t]{6cm}
\begin{center}
\includegraphics[width=6cm,clip]{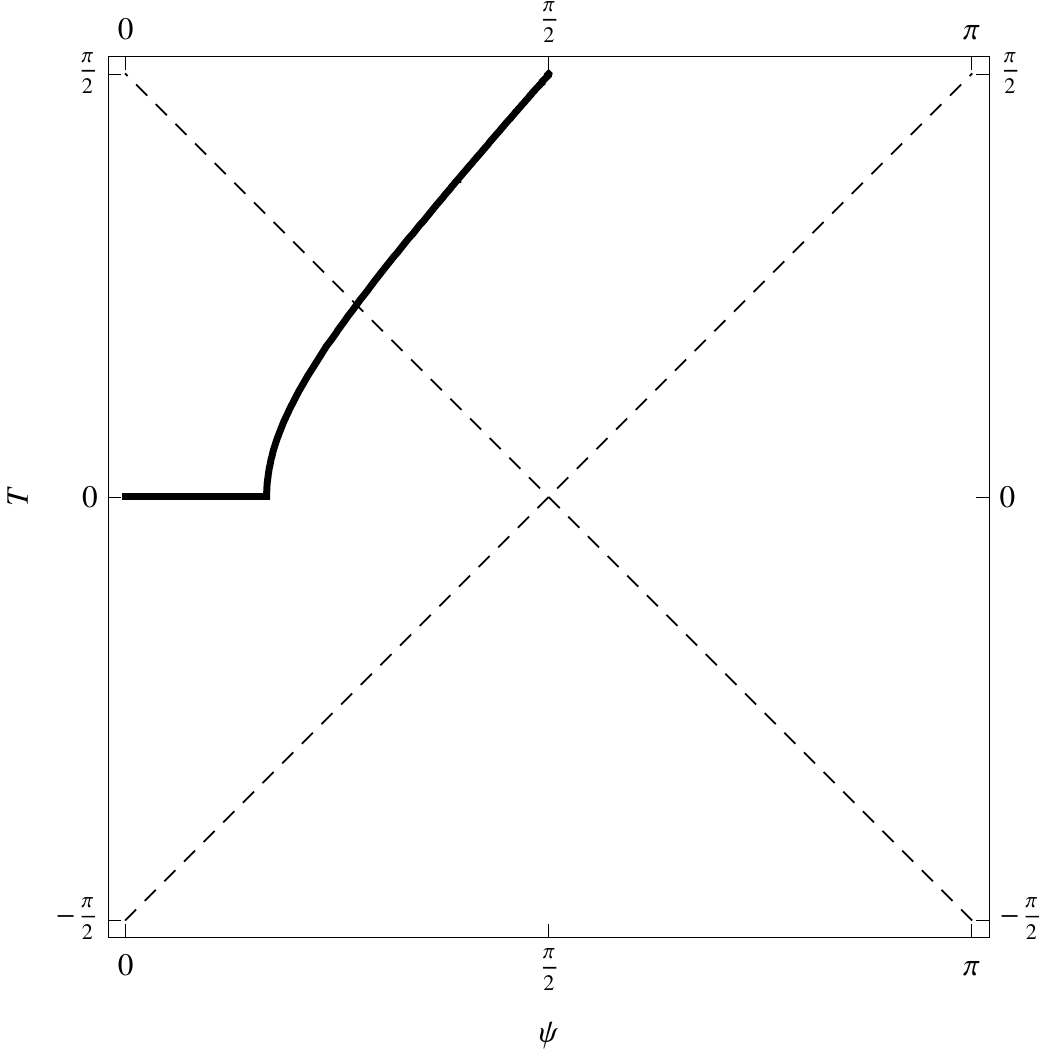}
\caption[Short caption for figure 1]{\label{subhor} Subhorizon bubble (conformal diagram for false vacuum observer)}
\end{center}
\end{minipage}
\hfill
\begin{minipage}[t]{6cm}
\begin{center}
\includegraphics[width=6cm,clip]{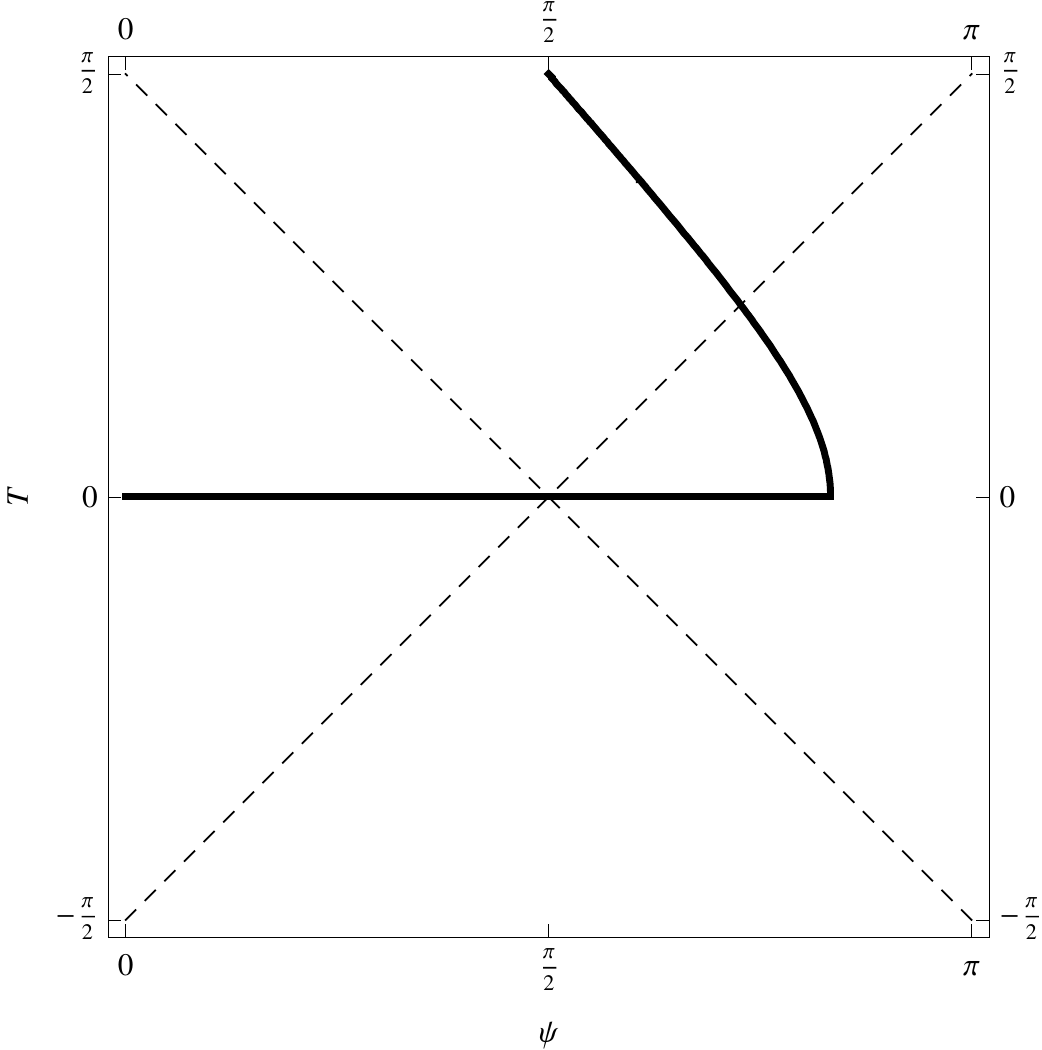}
\caption[Short caption for figure 2]{\label{superhor}Superhorizon bubble (conformal diagram for false vacuum observer)}
\end{center}
\end{minipage}
\end{figure}

For positive tension instantons, corresponding to the inner surfaces in Figure \ref{circles}, the boundary wall is located at
\be
x_1=R_{F} \sin(\omega/R_{F}) \cos(\psi) = \pm \sqrt{\rf^2-\rho^2}
\label{euctraj}
\ee
with $+$ for Type (a) and $-$ for Type (b).  We analytically continue $\omega \to i \eta + \frac{\pi}{2}$, giving us Lorentzian de Sitter metrics
\be
\td s^2=-\td\eta^2 + R_{(F,T)}^2 \cosh ^2(\eta/R_{(F,T)}) \td\Omega_{(3)}^2,\label{global}
\ee
in the true and false vacua, and 
giving the trajectory of the bubble wall in false-vacuum coordinates
\be
x_1=R_{F} \cosh(\eta/R_{F}) \cos(\psi) = \pm \sqrt{\rf^2-\rho^2}
\ee
where Types (a) and (b) again take opposite sign.  The conformal structure of these solutions is apparent if we define $\tau=2 \arctan(e^{\eta/R_{dS}})-\frac{\pi}{2}$, in which case (\ref{global}) becomes
\be
\td s^2=\sec(\tau) (-\td\tau^2+\td\Omega_{(3)}^2)
\ee
In these coordinates the trajectory of the bubble according to the false vacuum observer is 
\be
\sec(\tau) \cos(\psi) = \pm\sqrt{\rf^2-\rho^2}.
\ee
with Type (a) taking the positive sign and Type (b) the negative sign.  The corresponding Penrose diagrams are given in Figures \ref{subhor} and \ref{superhor}.  Both bubble trajectories asymptote to the horizon radius $\psi=\pi /2$ at future conformal infinity $\tau=\pi /2$.  However, for Type (a) instantons the bubble nucleates below the cosmological horizon and expands to fill the horizon. Type (b) solutions follow an apparently acausal trajectory, nucleating at a super-horizon size, and then shrinking back the horizon.  

The new negative-tension instantons follow a similar pattern.  They correspond to the outer regions in Figure \ref{circles}, which we can study by taking $x_1 \rightarrow -x_1$ in our previous formulae.  The false vacuum observer sees the bubble wall trajectory
\be
\sec(\tau) \cos(\psi) = \mp \sqrt{\rf^2-\rho^2},
\ee
where now the $(-)$ sign corresponds to Type (c) and the $(+)$ sign to Type (d).  These are the same trajectories as the positive-tension bubbles, but now Type (c) has the apparently acausal trajectory and Type (d) the causal one.

\begin{figure}
\begin{center}
\includegraphics[scale=.4]{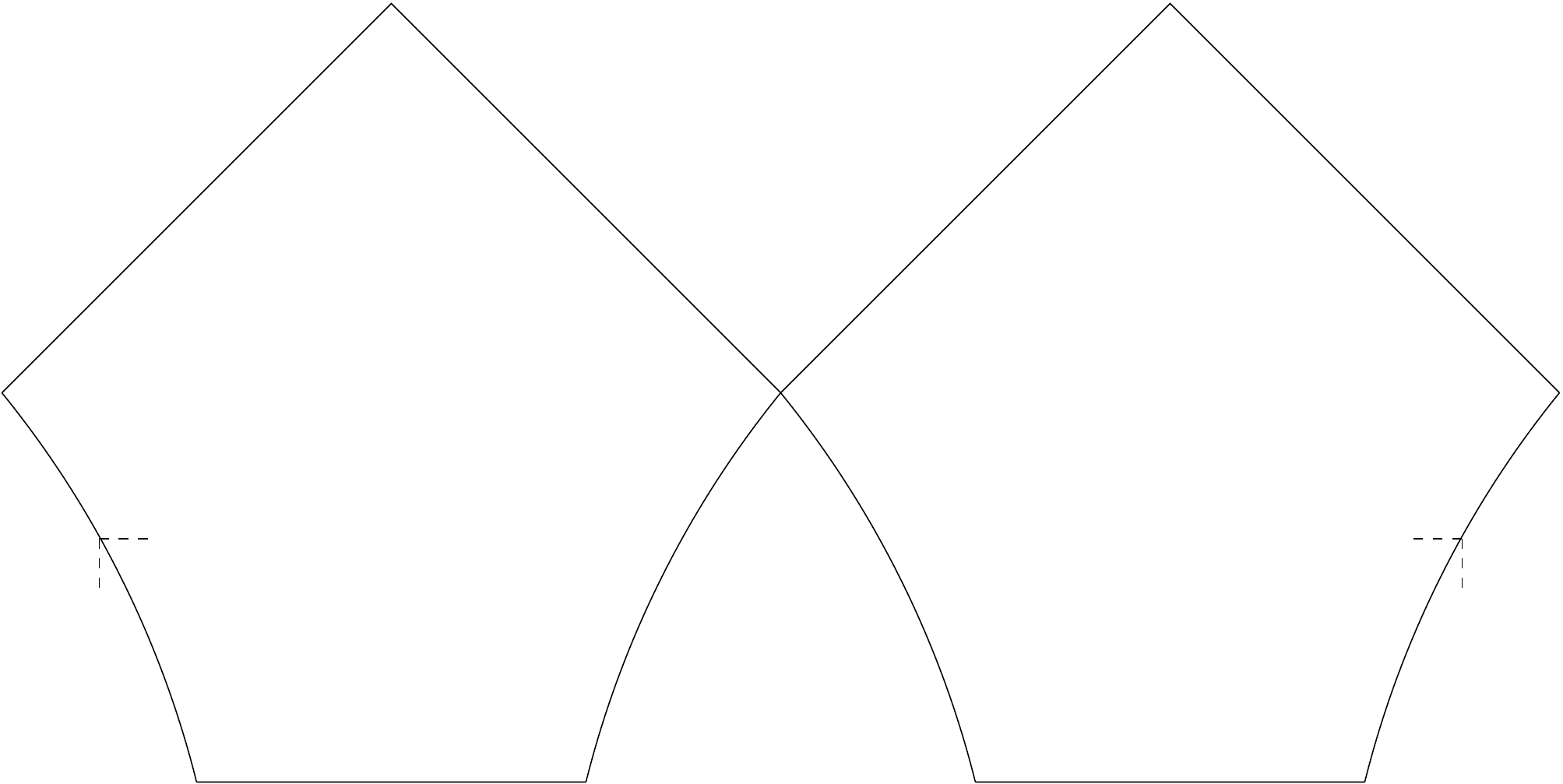}
\caption{Conformal diagram for Minkowski space ``bubble of nothing".  This is the ``upstairs" picture.  The $Z_2$ quotient is obtained by taking one half of this diagram.}
\label{flatflat}
\end{center}
\end{figure}

The final conformal diagram we need is the one for the Minkowski instanton described in Section \ref{ss:Z2Mink}.  The bubble wall trajectory in static coordinates is 
\be
r(t) = R \sqrt{1+\left( \frac{\rho^2}{R^2}-1\right) \mathrm{sech}^2(\frac{t}{R})}
\ee
and in the Minkowski space limit where the radius of curvature $R$ is taken to infinity, the trajectory becomes 
\be
r(t)=\sqrt{\rho^2+t^2}. \label{minktraj}
\ee
To interpret this result, we adopt light cone coordinates 
\begin{subequations}
\begin{align}
u&=t-r\\
v&=t+r
\end{align}
\end{subequations}
in which the flat metric is 
\be
\td s^2 = -\td u \td v+\frac{1}{4}(v-u)^2 \td\Omega_{(2)}^2.
\ee
With coordinates $u=\tan(p), v=\tan(q),$ the metric is 
\be
\td s^2=f(p,q)^2 (-\td p \td q +\frac{1}{4}\sin^2(p-q)\td\Omega_{(2)}^2 )
\ee
with $f$ a conformal factor.  Removing this conformal factor, and after the redefinition $p=\tau+\mathcal{R},q=\tau-\mathcal{R}$ the metric becomes 
\be
\td s^2=-\td T^2+d\mathcal{R}^2 +\frac{1}{4}\sin^2(2\mathcal{R}) \td\Omega_{(2)}^2
\ee
with $-\frac{\pi}{2} < \tau < \frac{\pi}{2}$ and $0<\mathcal{R}<\pi$.  In these coordinates the bubble wall trajectory is
\be
\tan(\tau+\mathcal{R}) \tan(\tau-\mathcal{R})=-\rho^2
\ee
as is illustrated in Figure \ref{flatflat}.

\subsection{Negative mode and causality}\label{ss:CausalityNegModes}

The conformal structures in the previous section describe bubbles that are both causal and acausal.  Solutions of Types (b) and (c) describe bubbles which form at superhorizon size, an apparent violation of causality.  The solutions of Types (a) and (d) describe causal bubbles which nucleate at subhorizon radius.

There is an interesting pattern here which can be revealed by inspection of Table \ref{sumtable}.  Namely, the causal instantons (a) and (d) possess negative modes, and so can be interpreted as tunneling solutions.  On the other hand, the acausal instantons (b) and (c) have no negative modes, and therefore should not be interpreted as tunneling solutions.  Though the requirement of a negative mode appears in some of the earliest work on CdL-type instantons, the interpretation and role of the negative mode has remained somewhat controversial \cite{Coleman:1987rm}-\cite{ColemanSteinhardt}.  We see here that by requiring the negative mode for a tunneling solution, we obtain a complete and self-consistent story: the negative mode requirement automatically restricts us to causal tunneling solutions.  Although 
we have not \emph{a priori} imposed causality restrictions, it seems that semiclassical tunneling is self-regulating in this respect.  This is further evidence that the negative mode requirement is justified for semiclassical tunneling solutions.

Physically, behavior of the negative mode is driven by a sort of ``enhanced symmetry point" (ESP) at the horizon. As the bubble nucleation radius approaches the horizon radius from below, the eigenvalue $\lambda_\rho$ switches from negative to positive, passing through zero when the nucleation radius equals the horizon radius.  At the horizon radius itself, $\lambda_\rho = 0$ and small fluctuations in the bubble nucleation radius have zero action.  At this point, the action has an extra symmetry corresponding to infinitesimal changes in the bubble nucleation radius.  Since $\lambda_\rho$ is vanishing, and since $\Gamma$ is proportional to $\lambda_\rho^{-1/2}$, at this point of enhanced symmetry the semiclassical tunneling rate diverges.  This indicates that the semiclassical tunneling prescription is breaking down for nearly horizon-sized bubbles.  Superficially, this divergence is similar to the ones which occur in effective field theory if one attempts to integrate out a massless degree of freedom.  Far away from the ESP the semiclassical tunneling rate is reliable, and in this regime the negative mode prescription automatically protects causality for all of the thin-wall instantons described here.

\section{Conclusions}\label{s:conclusions}

In this paper, we have generalized the familiar Coleman-DeLuccia tunneling process to include configurations in which the bubble wall may carry negative tension.  Provided the negative tension is sufficiently large in magnitude, these instantons satisfy the requirements for a tunneling solution.  They are solutions to the Euclidean equations of motion, and have a negative mode for fluctuations around the background.  We have provided two concrete realizations of these solutions in situtations where one might expect a negative tension object to arise.  Although these objects violate the null energy condition, and hence may be manifestations of the associated pathologies, they are automatically protected against causality violation in the same manner as positive tension bubble solutions.  

There are some interesting open questions concerning these negative-tension solutions.  In our ghost condensate construction in Section \ref{s:ghost}, we found a negative-tension solution that tunneled between two Minkowski vacua.  The scalar field gradient was everywhere spacelike.  One application of ghost condensates is in constructing de Sitter vacua with nontrivial properties, using scalar fields with timelike gradients. It would be interesting if a full CdL-like solution interpolating between two such de Sitter vacua could be found.  In the example we studied, we could not interpolate between vacua with a spacelike and timelike scalar field gradient because they were separated by a regime in which ghosts were present.  Finding a solution which could interpolate between timelike and spacelike field gradients would be helpful in connecting with cosmological applications of ghost condensates.

To construct a $Z_2$ symmetric boundary of spacetime, we used a simple model in which the orbifolding process localized negative stress-energy on the boundary surface.  This is consistent with many cosmological constructions that rely on negative tension, such as the Randall-Sundrum model.  We found that there is a negative mode, and hence a tunneling solution, provided the tension on the spacetime boundary is negative definite.  The orientifold planes of string and M theories possess these properties, but carry conserved charges related to their tensions.  Because of these conserved charges, simple generalizations of the instantons described here are forbidden by charge conservation.  There remains the possibility that these boundaries may be nucleated in some more complex process analogous to $e^\pm$ pair production, in which the configuration has zero net charge.  This would give a scenario similar to that proposed by Brown and Teitelboim \cite{Brown:1987dd,Brown:1988kg}, in which a flux background is unstable to the nucleation of such objects.  These would be interesting stringy backgrounds in their own right, and because they mediate instabilities they may serve to constrain models which include orientifold planes.

The solutions we have described have some puzzling features, but if one admits the reality of some of the ingredients which go into their construction (such as higher-derivative field theories or negative-tension boundaries) then we have argued that they should be possible.  Many ideas in higher-dimensional physics and cosmology suggest we include just such ingredients.  Since we have focused on some simple examples, the examples of instabilities given here are not sufficient to rule out any existing theories.  Nonetheless, since theories which include these exotic elements could suffer instablities similar to those we describe here, our solutions provide a new consistency constraint on these models.

\section*{Acknowledgements}

We are grateful to Adam Brown, Steven Gratton, Jean-Luc Lehners, Malcolm Perry, Andrew Tolley and Mark Wyman for comments.
\appendix
\section{Appendix: negative modes and gravity \label{appendix}}
In this paper we are concerned not with corrections to the ground-state energy, but with solutions which can be interpreted as tunneling events.  We therefore look for Euclidean-time solutions which correspond to unstable states.  These tunneling solutions must therefore satisfy two requirements: to be classed as instantons, they must satisfy appropriate boundary conditions and have finite Euclidean action, and to be describe tunneling, they must have a negative mode.

But a negative mode of what?  The generalization of the operator $\frac{\partial^2 S}{\partial \phi^2}$ to the gravitational case is not immediately obvious.  The analog of the CdL instanton solution $\phi_{cl}(x^{\mu})$ is now a metric on the instanton manifold, which we will denote $g_{\mu \nu}^{cl}$.  The exponential dependence of the tunneling rate $\Gamma /V$ is therefore $\exp(-B)$, where B  is given by $B=S_E(g_{\mu \nu}^{cl})-S_{Stay}$, and $S_{Stay}$ is the Euclidean action \emph{not} to tunnel; that is, of the surrounding false vacuum spacetime.  To find the prefactor, we consider small perturbations about the instanton metric 
$$g_{\mu \nu}=g_{\mu \nu}^{cl}+h_{\mu \nu}.$$
A general metric perturbation may be decomposed as 
\be
h_{\mu \nu}=h_{\mu \nu}^{(2)}+h_{\mu \nu}^{(1)}+h_{\mu \nu}^{(0)} +\frac{1}{4} g_{\mu \nu} h
\ee
with
\begin{eqnarray}
h_{\mu \nu}^{(0)}&=&(\nabla_{\mu}\nabla_{\nu}-\frac{1}{4} g_{\mu \nu} \nabla_{\sigma}\nabla^{\sigma} )C \label{hT} \\
h_{\mu \nu}^{(1)}&=&\nabla_{\mu} \eta_{\nu}+\nabla_{\nu} \eta_{\mu} \label{hL}\\
\nabla_{\mu} h_{\mu \nu}^{(2)}&=& h^{\nu}_{\nu}=0 \label{hTT}\\
\end{eqnarray}
where $C$ is a scalar, and $B$ is a divergence-free vector $ \nabla^{\mu}\eta_{\mu}=0$.  Following Gibbons and Perry \cite{Gibbons:1978ji}, we define the generalized Laplacian operators 
\begin{eqnarray}
\Delta_{(0)} \phi^{(n)}&=&-\nabla_{\sigma}\nabla^{\sigma} \phi^n \label{Laplace:scalar}\\
\Delta_{(1)} \phi^{(n)}_{\mu}&=&-\nabla_{\sigma}\nabla^{\sigma}\phi_{\mu}^n-2 R_{\mu \nu} \phi^{(n)\nu}\label{Laplace:vector}\\
\Delta_{(2)} \phi^{(n)}_{\mu \nu}&=&-\nabla_{\sigma}\nabla^{\sigma}\phi^n_{\mu \nu}-2 R_{\mu \nu \rho \sigma}\phi^{(n)\nu \sigma}\label{Laplace:tensor}
\end{eqnarray}
where $\Delta_{(0)}$, $\Delta_{(1)}$, and $\Delta_{(2)}$ act on scalars, divergence-free vectors, and symmetric, transverse, traceless tensors respectively.    Expanding the action to second order about the instanton solution will therefore produce a term quadratic in the metric fluctuations $h$ which involves these operators.  

By generalized coordinate invariance, we are free to choose a gauge fixing condition $F^{\mu},$ which we assume takes the linearized form 
\be
F^{\mu}=\sqrt{-g} \left(\nabla^{\nu} h^{\mu}_{nu} - \frac{\beta}{2} \nabla^{\mu} h^{\nu}_{\nu}\right)\label{lineargauge}
\ee
where $\beta$ is the gauge parameter.  This choice can be implemented in a gravitational path integral by adding the Faddeev-Popov ghost Lagrangian 
$$\mathcal{L}_{ghost}=\bar{C}_{\mu} \left( \frac{\delta F^{\mu}}{\delta \xi^{\nu}}\right)C^{\nu}$$
where $\delta \xi^{\nu}$ parametrizes the coordinate transform.  The functional integral therefore contains a Faddeev-Popov determinant in addition to the terms stemming from the Laplacians in (\ref{Laplace:scalar} - \ref{Laplace:tensor}).  

We will not attempt to construct an explicit expression for the term quadratic in the metric perturbation; a full calculation can be found in Refs. \cite{Allen:1986tt,Gibbons:1978ji,Yasuda:1983hk}.  The end result is surprisingly simple.  For the simple class of linear gauges (\ref{lineargauge}), the determinant prefactor is gauge-independent, and is
\be
A\sim(\det \Delta_{(1)})^{1/2}(\det \Delta_{(2)})^{-1/2} \label{gravdet}
\ee
The contributions of the scalar operator $\Delta_{(0)}$ are cancelled by the Faddeev-Popov determinant which arises from incorporating the gauge-fixing condition \eqref{lineargauge}. 

Incorporating the gauge-fixing condition (\ref{lineargauge}) does not, however, completely eliminate the problem of zero modes.  Gauge-fixing cancels the zero-mode contributions which arise from generalized coordinate invariance \cite{Allen:1986tt, Yasuda:1983hk}.  But there can be more symmetries in the problem.  Suppose the instanton metric $g_{\mu \nu}^{cl}$ is invariant under a $d_i$-dimensional isometry group.  Then we can define $d_i$ Killing vectors, which must satisfy
\be
\nabla_{(\mu}K_{\nu)}=0.\label{Killing}
\ee
Applying $-\nabla_{\mu}$ to (\ref{Killing}), we find 
$$-\nabla_{\mu} \nabla^{\mu} K_{\nu} -2 R_{\mu \nu} K^{\nu}=0$$
and the Killing vectors live in the kernel of the operator $\Delta_{(1)}$.  We can handle these zero modes in a similar manner to the ones arising from coordinate translations in the scalar field space, and define 
$\det^{\prime}\Delta_{(1)}$ to be the determinant with these modes projected out. 
 Moreover, the identity \cite{Gibbons:1978ji}
$$\int_{M} \sqrt{-g}\left( \nabla_a V^{a} \nabla_b V^b - 2 \nabla_{(a}V^{b)} \nabla_{(b}V^{a)}+V_a\Delta_{(1)}V^a \right)\; d^4 x=0,$$
which follows from the definition of $\Delta_{(1)}$, holds for any divergence-free vector $V$.  This establishes that this operator is positive semi-definite, and contains no negative modes. 

The zero modes of $\Delta_{(2)}$ arise from transverse, traceless metric perturbations constructed from physical symmetries of the problem, and the negative modes of this operator reveal the instability of a gravitational vacuum.
\subsection{Instabilities and the boundary surface}
As we showed in Chapter 2, the operator $\mathcal{D}_E$ can be decomposed into components which act on scalars, vectors, and tensors.  Negative modes are associated only with the tensor component.  To look for instabilities, we therefore want to investigate the spectrum of the second-derivative operator acting on transverse, traceless, symmetric metric perturbations $\phi^{ab}$:
\be
\triangle_{(2)}=-\nabla^c \nabla_c \phi_{ab} - 2 R_{acbd} \phi^{cd} = \lambda \phi_{ab} \label{e:gop}
\ee
The complete spectrum for the bounce metric remains an open question, but fortunately we are interested only in the existence of an $O(4)$-symmetric negative mode.  In this section we will show that negative modes, if they exist, must have nonzero components in the direction normal to the join surface.  

The operator \eqref{e:gop} has no negative eigenvalues if and only if the B\"ochner integral over the entire manifold is positive:
$$
\int _{\mathcal{M}}\sqrt{g}  d^4 x ( -\phi_{ab} \nabla^c \nabla_c \phi^{ab}-2 R_{acbd} \phi^{ab} \phi^{cd}) .
$$
The wave operator $- \nabla^c \nabla_c$ is positive definite on a closed manifold provided its action is restricted to TT tensors.  This is easily shown by integrating the operator by parts over a manifold without boundary.  The CdL instanton, however, is constructed from two four-spheres joined along a three-dimensional surface; the manifold therefore has a natural boundary over which the gradient operator $\nabla$ is discontinuous.  We therefore have
$$-\int _{M} \sqrt{g} d^4 x( - \phi_{ab} \nabla^c \nabla_c \phi^{ab}) = \int _{M}  \sqrt{g} d^4 x(\nabla_c \phi^{ab})^2  - \int_{\partial M}n_{c}(\phi_{ab} \nabla^{c} \phi^{ab})d^3x  $$
where $n$ is a vector normal to the join surface.  The existence of a boundary may therefore induce a negative mode in the wave operator defined on the instanton manifold.  

We now turn our attention to the dependence of the operator on the Riemann tensor $R$. It is useful to define a four-dimensional Riemann tensor for the instanton manifold $\widetilde{M}$ via the Gauss-Codazzi equation
\be
^{(4)}\tilde{R}_{acbd} =^{(3)}R_{acbd} + \tilde{K}_{ab} \tilde{K}_{cd} - \tilde{K}_{bc} \tilde{K}_{ad},
\ee
where the extrinsic curvature is taken to be the average $\tilde{K}_{ab} = \frac{1}{2} \left( K_{ab}^F + K_{ab}^T\right)$.  We can now use the maximal symmetry of the boundary $S^3$ to find a simple expression for the Riemann tensor on the buble wall 
$$^{(3)}R_{acbd}=\frac{1}{b^2(\sigma_{wall})}(h_{ab} h_{cd}-h_{ad}h_{bc}).$$
Using (\ref{exdef}), we can now write
$$^4R_{acbd} \phi^{ab} \phi^{cd} = (b^{-2} +\widetilde{\left(\frac{b'}{b}\right)}^2)(h_{ac} h_{bd}-h_{ad}h_{cb})\phi^{ab} \phi^{cd}$$ 
where $$\widetilde{\frac{b'}{b}}=\frac{1}{2}\left(\frac{ b'_T}{b_T} + \frac{b'_F} {b_F}\right).$$

Because $\phi_{ab}$ is traceless, we have
$$-2\widetilde{R}_{acbd} \phi^{ab} \phi^{cd}  = 2\left\{b^{-2} +\widetilde{\left(\frac{b'}{b}\right)}^2\right\}(\phi_{ab} \phi^{ab} -2 n_a n_d \phi^{a}_c \phi^{cd}).$$
The term in curly brackets is positive definite, so once again we find that there must be some component  of the perturbation in the normal direction in order to induce a negative mode. 

We see that if a perturbation to the instanton metric \eqref{EucMetric} is to be an eigenfunction of $\Delta_{(2)}$ with a negative eigenvalue, it cannot be ``pure bulk" and must have some components defined on the boundary.  The perturbation corresponding to the lowest eigenvalue must satisfy
\begin{eqnarray}
n^c \phi_{ab} \nabla_c \phi^{ab}> 0 \nonumber \\
n_a n^b \phi^{ac}\phi_{cb} > 0 \label{bconds} .
\end{eqnarray}

We will now show that there is no $O(4)$-symmetric perturbation mode defined in the bulk, and that any negative mode which preserves the symmetries of the instanton must be confined to the boundary.  The most general traceless $O(4)$-symmetric perturbation must take the form
\begin{eqnarray}
\phi_{00}&=&A(\sigma)\nonumber \\
\phi_{0i}&=&0\nonumber \\
\phi_{ij}&=&-\frac{1}{3} A(\sigma)\delta_{ij} \label{ea:TTpert}
\end{eqnarray}
We now demand that this be transverse in order to determine the function $A(\sigma)$.  Using the covariant derivative defined with respect to the instanton metric, we find the condition 
$$\nabla_{a}\phi^{ab} =0 \implies A(\sigma) = \frac{C}{b^4(\sigma)}$$
with $C$ an arbitrary constant.  In order for the instanton manifold to satisfy the appropriate boundary conditions, we must have $b(\sigma) \sim 0$ at the north and south poles of the Euclidean manifold.  The perturbation \eqref{ea:TTpert} is therefore singular at $\sigma = 0$ and $\sigma = \pi \rt$, and the Euclidean action for the perturbation is not finite.   Perturbations of this form are therefore excluded from the path integral and do not represent eigenfunctions of the operator $\Delta_{(2)}$.  This is commensurate with the fact that there are no $l=0$ (or indeed $l=1$) tensor eigenfunctions for pure de Sitter.

We have shown that any negative mode must have nonzero boundary components  \eqref{bconds}, and that there is no $O(4)$-symmetric mode in the bulk. Therefore the only modes which preserve the symmetries of the instanton solution correspond to shifts in the boundary, as claimed in Section \ref{ss:NegModes}.

\end{document}